 \documentclass[12pt,a4paper]{elsarticle}

\usepackage{amsmath,amsthm}
\usepackage{mathtools}
\usepackage{graphicx}
\usepackage{tabularx}
\usepackage{floatpag}
\usepackage{float}
\usepackage{xcolor}
\usepackage[colorlinks=true,linkcolor=blue]{hyperref}
\usepackage{caption}
\usepackage{subcaption}
\usepackage{multirow}
\usepackage{natbib}
\usepackage{ulem}
\biboptions{numbers}

\usepackage{tikz,pgfplots} 
\usepackage{overpic}
\pgfplotsset{compat=1.3}
\usetikzlibrary{external}
\let\realverbatim=\verbatim
\let\realendverbatim=\endverbatim
\renewcommand\verbatim{\par\addvspace{6pt plus 2pt minus 1pt}\realverbatim}
\renewcommand\endverbatim{\realendverbatim\addvspace{6pt plus 2pt minus 1pt}}

\emergencystretch=\maxdimen
\newcommand{\red}[1]{\textcolor{red}{#1}}

\newcommand\Pe{\mbox{Pe}^{-1}}            

\newsavebox{\astrutbox}
\sbox{\astrutbox}{\rule[-5pt]{0pt}{20pt}}

\newcommand\p{\ensuremath{\partial}}

\newcommand{\ba}{\begin{align}}
\newcommand{\ea}{\end{align}}

\newcommand{\pad}[2]{\frac{\p #1}{\p #2}}
\newcommand{\padd}[2]{\frac{\p^2 #1}{\p {#2}^2}}

\newcommand\ra{\rightarrow}
\newcommand\Ra{\Rightarrow}

\newcommand{\Di}{K_{di}}
\newcommand{\Aii}{K_{ai}}

\journal{Partial Differential Equations in Applied Math.}

\begin{document}

\begin{frontmatter}

\title{Mathematical modelling of flow and adsorption in a gas chromatograph}

\author[1]{A. Cabrera-Codony}
\affiliation[1]{organization={LEQUIA, Institute of the Environment, Universitat de Girona}, 
           country={Spain}}
           
\author[2]{ A. Valverde}
\affiliation[2]{organization={Department of Chemical Engineering, Universitat Politècnica de Catalunya}, 
           country={Spain}}
           
\author[3]{ K. Born}
\affiliation[3]{organization={School of Computer Science and Applied Mathematics, University of the  Witwatersrand, Johannesburg}, 
           country={South Africa}} 
           
 \author[4]{ O.A.I. Noreldin}
\affiliation[4]{organization={ Department of Mathematical Sciences,  University of Zululand}, 
           country={South Africa}} 
 \author[5]{ T.G. Myers\thanks{Corresponding Author: {\tt tmyers@crm.cat}}}
\affiliation[5]{organization={Centre de Recerca Matemàtica},
           state={Barcelona},
           country={Spain}}

\begin{abstract}
In this paper, a mathematical model is developed to describe the evolution of the concentration of compounds through a gas chromatography column. The model couples mass balances and kinetic equations for all components. Both single and multiple-component cases are considered with constant or variable velocity. Non-dimensionalisation indicates the small effect of diffusion. The system where diffusion is neglected is analysed using Laplace transforms. In the multiple-component case, it is demonstrated that the competition between the compounds is negligible and the equations may be decoupled. This reduces the problem to solving a single integral equation to determine the concentration profile for all components (since they are scaled versions of each other). For a given analyte, we then only two parameters need to be fitted to the data. To verify this approach, the full governing equations are also solved numerically using the finite difference method and a global adaptive quadrature method to integrate the Laplace transformation. Comparison with the Laplace solution verifies the high degree of accuracy of the simpler Laplace form. The Laplace solution is then verified against experimental data from BTEX chromatography. This novel method, which involves solving a single equation and fitting parameters in pairs for individual components, is highly efficient. It is significantly faster and simpler than the full numerical solution and avoids the computationally expensive methods that would normally be used to fit all curves at the same time.
\end{abstract}




\end{frontmatter}




\section{Introduction}\label{sec:intro} 

Gas chromatography (GC) is a widely used technique for identifying and analysing volatile compounds. It has a broad range of applications, such as detecting and quantifying pollutants, pesticides, and environmental contaminants in air, water, and soil samples \cite{boada2021core,cabrera2021biocollagenic}. In food and beverage analysis, it is used to determine the presence and concentrations of flavour compounds or additives. In the pharmaceutical industry, GC is employed for analysing drugs, including their purity, and the quantification of active ingredients. Additionally, clinical and medical laboratories use GC for analysing blood, urine, and other biological samples \cite{vspanik2018recent,zhou2022chemical}.

GC operates through a series of precise steps to separate and analyse volatile compounds within a sample. Initially, a small quantity of the sample is injected into the chromatograph. A carrier gas, typically helium or nitrogen, transports the vaporised sample through a chromatographic column. The column may be lined with a thin layer of liquid or packed with adsorbent material. The flowing gas is termed the mobile phase, while the lining or packed solid adsorbent is called the stationary phase- nowadays most GC systems involve a polymeric liquid stationary phase. Typical dimensions for a column are of the order 5-50m long with an inner diameter of 100–300 microns. The process's temperature is controlled by placing the column inside an oven. 

The separation process is caused by the interaction between the sample compounds and the stationary phase. As the sample travels through the column, the compound's molecules are repeatedly adsorbed and desorbed by the stationary phase. The rates of attachment differ for different molecules (depending on, for example, their molecular size, polarity, and volatility) and different types of stationary phases. Compounds that interact strongly with the stationary phase spend more time in the column and move slower, while those with weaker interactions move through quicker. Oven temperature programming is often employed, where the column temperature is gradually increased during the process. This helps achieve efficient separation by progressively reducing the compounds' affinity for the stationary phase, allowing them to elute based on their boiling points and interactions with the stationary phase. Upon exiting the column, the mixture passes through a detector which identifies the separate compounds. Common types of detectors used in GC include flame ionisation detectors, electron capture detectors, and mass spectrometers. The signals produced by the detector are then recorded and analysed to create a chromatogram. The chromatogram is a graphical representation that shows the concentration of compounds over time. Typically it takes the form of a series of Gaussian-like and asymmetric peaks  \cite{GrobBarry}, which appear at different times. The shape of the elution peak and retention time are affected by operating parameters and materials.

Various authors have carried out theoretical and experimental studies on liquid chromatography, as documented in \cite{GrobBarry, aldaeus2007prediction,degerman2007modeling, kaczmarski2020note,karolat2010prediction}. The importance of the retention time and peak width in gas chromatography columns has been addressed in studies by Dose \cite{dose1987simulation} and  Rodríguez \textit{et al}. \cite{cuevas2021numerical}.
Laplace transforms have proved effective in the study of single components, under a number of restrictions.  Guiochon and  Lin \cite{Guiochon} present a solution for a pulse inlet condition after neglecting diffusion. Noting that the inverse transform is complicated they reduce this condition to a delta function. They also discuss the case with diffusion but are forced to apply a constant concentration at the inlet. Their solution follows the work of Lapidus and Amundson \cite{Lapi52} which is valid for a semi-infinite column. In all cases, the velocity remains constant throughout the column and only a single component is analysed.

Mathematical models for chromatography are separated based on the type of adsorption and operating parameters. Linear equilibrium elution in GC with capillary columns has been investigated by 
Aris \cite{ARIS1959} and Golay\cite{Golay1958}. Packed column GC is well studied across linear/nonlinear equilibrium/nonequilibrium elutions \cite{KUCERA1965,GRUBNER1967,YAMAOKA1974,Romdhane1993,Lee1988, VIDAL1977} and similarly with open columns \cite{Pawlisch1987, Jaulmes1984}. More recently the work done by Guiochon and Lin \cite{Guiochon} and Asnin \textit{et al.} \cite{ASNIN2007} in which nonlinear isotherms were considered, takes into account both the analyte concentrations in the liquid phase and sites of adsorption in the stationary phase. 

Rodríguez \textit{et al.} \cite{cuevas2021numerical} proposed a mathematical model based on the advection-diffusion equation and Langmuir kinetic equation to analyse the transport of BTEX (benzene, toluene, ethylbenzene, and xylene isomers) molecules inside a capillary chromatography column. This model was solved numerically using a  finite volume method, and the results were validated with experimental data on  BTEX compounds. This study highlighted the importance of considering proper volatile organic compounds (VOCs) separation for accurate quantification and monitoring of pollutants.

A similar system but with  a linear kinetic equation was employed to investigate liquid chromatography in a packed fixed cylindrical column \cite{ahmad2023numerical,perveen2022simulations}. In both studies, the finite volume method was utilised to obtain numerically solutions, analyse the performance of the underlying process, investigate retention behavior, and identify the optimal parameter values in the liquid chromatography process. Perveen et. al. \cite{perveen2022simulations} employed  a linear kinetic equation to describe the rate of change of concentration. For the computations they employed the so-called bi-Langmuir isotherms, which are inconsistent with the kinetic equation (the isotherm comes from the steady-state of the kinetic equation).

Mathematical models for GC are analogous to the capture of contaminants in a packed columm. The theory in this field for the single contaminant case is well established and has been recently advanced in
\cite{myers2020mass,myers2020mathematical,myers2023development}. These papers involve the coupling of an advection-diffusion equation describing the mass balance with linear and nonlinear kinetic equations. Analytical solutions are obtained through identifying dominant terms and applying a travelling wave substitution. Comparison with numerical solutions and experimental data confirm their accuracy, in particular showing significant improvements over standard models (many of which are presented in the review of \cite{patel2019fixed}). Aguareles\textit{ et al.} \cite{Agua23} extend the results to include chemical bonding, providing a family of solutions for different reactions. The difference between such models and previously accepted standard ones is clarified in \cite{Myers24}. In particular it is explained how the travelling wave solutions maintain accuracy over a wide range of operating conditions while earlier models tend to be accurate for a single experiment but fail when conditions change. 

The objective of the present study is to develop and analyse a mathematical model for GC. The model development will follow that of  \cite{myers2023development, Myers24} for column adsorption. The purpose of the study being to gain a deeper understanding of the behaviour of materials whilst passing through a column and so, once developed and verified, the model may be used as a tool for analysing and optimising the separation process in gas chromatography.

\section{Mathematical model}\label{mathmodel}

Here we present two basic models for the flow, adsorption and desorption of molecules in a long thin column. Assuming that the sample is introduced in a short burst at time $t=0$ we first consider the case where the velocity is constant. This is based on the fact that after a sufficiently large distance into the column, we expect the sample to have mixed with the carrier fluid, since the carrier fluid occupies a large volume compared to the sample the removal of small quantities should have a negligible effect on the flow. In the second case, we allow for density variation due to the pressure drop along the column such that the flow varies. In both cases, we assume that the stationary phase is always far from being fully loaded. This is in contrast to the packed column models of \cite{myers2020mass,myers2020mathematical,myers2023development}, where the intention is to retain as much contaminant as possible. With GC a small sample is introduced to a very long column and the desorption is at a similar rate to adsorption, consequently only partial loading occurs.

\subsection{Constant velocity model}

Under the assumption of a constant fluid velocity, $u$,  the evolution of the cross-sectional average concentration in the gas mixture and the amount captured by the stationary phase may be expressed by
\begin{align}
\pad c t + u \pad c x &= D \padd c x - \alpha \pad q t, \label{Eq1}
\\
\pad q t &= k_a c - k_d q \, .\label{Eq2}
\end{align}
Equation \eqref{Eq1} is a mass balance for the concentration density $c$, $D$ represents the diffusion coefficient, and $q$ is the amount attached to the stationary phase. 
The coefficient $\alpha=2 \delta/R$ is known as the phase ratio and represents the difference between the volume of stationary and mobile phases assuming the stationary phase coats a circular cylindrical tube ($\delta$ is the thickness of the stationary phase layer).
The derivation of \eqref{Eq1} is provided in \ref{App:constantu}.
Equation \eqref{Eq2} represents the attachment/detachment process, where $k_a, k_d$ are the adsorption and desorption coefficients respectively. It may be viewed as a reduced form of Langmuir kinetic equation where the amount attached at any moment is significantly lower than the attachment capacity (this follows from the assumption that the loading of the stationary phase is low in GC).

To determine the relative strength of the terms in the equations we non-dimensionalise the variables
$$
\hat x = \frac{x}{\mathcal{L}} \qquad \hat t = \frac{t}{\tau} \qquad \hat c = \frac{c}{c_0} \qquad \hat q = \frac{q}{q_e}
$$
where
$$
q_e=\mathcal{K} c_0 \qquad \mathcal{L}=\frac{u R}{2\delta k_a}   \qquad \tau = \frac{q_e}{k_a c_0} \, .
$$
The scale $q_e$ is the  adsorbed quantity in equilibrium. It is defined from the steady-state of \eqref{Eq2}, where $\mathcal{K}=k_{a}/k_{d}$ and $c_0$ the concentration at the inlet.
The time-scale $\tau$ indicates the order of magnitude of the time taken for the attachment process, that is we work on the attachment time-scale rather than the faster flow time-scale. The length-scale 
$\mathcal{L}$ is then the distance travelled by the fluid over the attachment time scale. 

The governing equations may now be written as 
\begin{align}
\text{Da}\pad{\hat c}{\hat t} +  \pad{\hat c}{\hat x} &= \Pe \padd{\hat c}{\hat x} -  \pad{\hat q}{\hat t} 
\\
\pad{\hat q}{\hat t} &= \hat c -  \hat q
\end{align}
where $\text{Da}=\mathcal{L}/(u\tau)$ is the Damk\"{o}hler number and $ \Pe = D/(u\mathcal{L})$ is the inverse P\'eclet number. With hydrogen as the carrier gas, moving at 2cm/s in a column with dimension 100$\mu$m the Reynolds number is of the order 10$^{-2}$ and the flow is clearly laminar, diffusion then is purely the result of Brownian motion. Noting that Brownian diffusion is generally negligible in comparison to advection the P\'eclet number term may be neglected \cite{Agua23,myers2023development}.  
The Damk\"{o}hler number $\text{Da}$ is also expected to be small however close to the start of the process there will be a time boundary layer where this term may be important, particularly for the numerical scheme which requires the term to be retained in order to apply the initial condition.  Consequently, for a single compound we define the  base system 
\begin{align}
\label{BaseC}
\text{Da}\pad{\hat c}{\hat t} +  \pad{\hat c}{\hat x} &= -  \pad{\hat q}{\hat t} 
\\
\label{BaseQ}
\pad{\hat q}{\hat t} &= \hat c - \hat q \, .
\end{align}

Assuming the column is initially free of the sample, which is introduced at the inlet over a short period, we apply
\begin{align}
    \hat c(\hat x,0)= q(\hat x,0) = 0 \qquad \hat c(0,\hat t) = H(\hat t)-H(\hat t-\hat t_1),
\end{align}
where $H$ represents the Heaviside function and the sample is injected for $ 0 \le \hat t \le \hat t_1$.

Since the stationary phase is far from equilibrium we can assume that
there is no competition for attachment sites (since many are available) and then the extension to an arbitrary number of components is trivial
\begin{align}
\text{Da}\pad {\hat c_i}{\hat t} +  \pad {\hat c_i}{ \hat x} &=  - \beta_i \pad {\hat q_i} {\hat t}  \, , 
\qquad \pad {\hat q_i} {\hat t}  = K_{ai}\hat c_i - K_{di} \hat q_i
\label{ArbEqs}
\end{align}
with $i = 1, ... , N$. The concentrations and adsorbed fractions are scaled with the inlet and equilibrium values of each component,  $\hat c_i=c_i/c_{0,i}$, $\hat q_i=q_i/q_{e,i}$ and
$$
q_{e,i}=\mathcal{K}_{i}c_{0,i} \,, \quad \mathcal{L}=\frac{u_0 R}{2\delta k_{a,1}}\,, \quad \tau=\frac{q_{e,1}}{k_{a,1}c_{0,1}}\, ,
$$
where $\mathcal{K}_{i}=k_{a,i}/k_{d,i}$ and $u_0=u$ is the inlet velocity.
The additional parameters arise due to the choice of time and length scales being based on component $i=1$ (which must then be chosen as a dominant component), such that 
$$
\beta_i=\frac{q_{e,i}c_{0,1}}{q_{e,1}c_{0,i}}\,, \quad \; K_{a,i}=\frac{k_{a,i}c_{0,i}q_{e,1}}{k_{a,1}c_{0,1}q_{e,i}}\,, \quad  \; K_{d,i}=\frac{k_{d,i}q_{e,1}}{k_{a,1}c_{0,1}}\,,
$$
and $\beta_1=K_{a,1}=K_{d,1}=1$.

In the absence of competition between compounds this multi-component model effectively reduces to a set of identical single component equations. In which case, it is sufficient to solve only a single pair of equations and then the appropriate solution for each component appears due to the different non-dimensionalisation. 

\subsection{Variable velocity model}\label{sec:modelvariableu}

Chromatography columns are very long compared to their inner diameter. To drive the flow then requires a significant pressure drop which may affect the gas density and thus  the velocity field of the flow.  This results in a modification to \eqref{Eq1}, such that 
\begin{align}
\pad c t + \pad{}{x} (uc) &=  \pad{}{x} \left(D \pad  c x\right) - \alpha \pad q t \, .\label{NotEq1}
\end{align}
Equation \eqref{Eq2} remains unchanged.
As noted in the studies of packed columns of \cite{myers2020mass, Myers24b} with large mass removal we must also track the motion of the carrier fluid. 
Anticipating the extension to multi-components, we denote the concentration of carrier gas molecules as $c_N$ and then 
\begin{align}
u=-\frac{R^2}{8\mu}\frac{\partial p}{\partial x} \, , \qquad \qquad
    p=p_0c_N/c_{0,N}\,  , \label{umultireduxDIM}
\end{align}
where $p_0, c_{0,N}$ denote the inlet values and $\mu$ the
dynamic viscosity. The carrier gas concentration also satisfies a mass balance of the form \eqref{NotEq1}, but with zero adsorption. 
Boundary and initial conditions follow those of the previous section.
The full derivation of the variable velocity model is detailed in \ref{App:variableu}. A similar variable flow model has previously been considered by Rodríguez \textit{et al}. \cite{cuevas2021numerical}.

If we distinguish the individual components by subsript $i$, where $i \in [1, N]$ represents different analytes and $i=N$ the carrier gas then the extension of the mass and momentum balance to multiple components is
\begin{align}
    &\frac{\partial  c_i}{\partial t}+ c_i\frac{\partial u}{\partial x}+ u\frac{\partial c_i}{\partial x}=\frac{\partial}{\partial x}\left(D_i\frac{\partial c_i}{\partial x}\right)-\alpha \frac{\partial  q_i}{\partial t} \quad \text{for} \quad i=1,...,N \, , \label{cmultireduxDIM}
    \\
     &\frac{\partial q_i}{\partial  t}=k_{a,i} c_i-k_{d,i} q_i\, \label{qmultireduxDIM} \, ,
    \end{align}
    and $q_N=0$.
The velocity and pressure for the carrier fluid are given by equations \eqref{umultireduxDIM}.

The initial and boundary conditions are
\begin{align} 
    &c_i(x,0)= q_i(x,0) = 0\,, \label{advdiffICu}\\
    &\left(uc_i-D_i\frac{\partial c_i}{\partial x}\right)\Bigg\vert_{x=0^+} = u_0c_{0,i}\left( H(t)-H(t- t_1)\right) \; \text{for}\; i=1,...,N-1 \,, \label{advdiffBCx0u}\\
    &\left(u c_N-D_N\frac{\partial c_N}{\partial x}\right)\Bigg\vert_{x=0^+} = u_0c_{0,N} \,,\label{advdiffBCx0Nu}\\
    &\frac{\partial c_i}{\partial x}\Bigg\vert_{x=L^-} = 0 \,, \qquad \qquad 
    p(L^-,t)=p_L \,,\label{advdiffBCxLu}
\end{align}
where $u_0 = u(0^-,t)$ is the fluid velocity just before the inlet, it may be calculated from the inlet mass flow, and $p_L$ is the pressure just before the outlet $x=L^-$.

The variables are nondimensionalised with 
$$
\hat x = \frac{x}{\mathcal{L}}\,, \quad \hat t = \frac{t}{\tau}\,, \quad \hat c_i = \frac{c_i}{c_{0,i}}\,, \quad \hat q_i = \frac{q_i}{q_{e,i}}\,, \quad \hat u = \frac{u}{u_0}\,, \quad \hat{p}=\frac{p}{p_0} \, ,
$$
where the scales match those of the previous section.

Again assuming $\text{Pe}_i^{-1}=D_i/(u_0\mathcal{L})\ll 1$, we may write the reduced dimensionless model
\begin{align}
    &\text{Da}\frac{\partial  c_i}{\partial\hat t}+\hat c_i\frac{\partial \hat u}{\partial \hat x}+\hat u\frac{\partial \hat c_i}{\partial \hat x}=-\beta_i\frac{\partial \hat q_i}{\partial \hat t} \quad \text{for} \quad i=1,...,N-1 \, , \label{cmultireduxND} \\
    &\hat c_N=1/\hat u \, , \label{cNmultireduxND} \\
    &\frac{\partial \hat q_i}{\partial \hat t}=K_{a,i} \hat c_i-K_{d,i}\hat q_i\,,\label{qmultireduxND} \\
    &\hat p=\sqrt{1-\left(1-\hat p_L^2\right)\hat x/\hat L}\,, \label{pmultireduxND}\\
    &\hat u=1/\hat p\,,\label{umultireduxND}
\end{align}
with 
\begin{align}
&\text{Da}=\frac{Rc_{0,1}}{2\delta q_{e,1}}\,, \quad  \hat p_L=\frac{p_L}{p_0} \,, \\
K_{a,i}=&\frac{k_{a,i}c_{0,i}q_{e,1}}{k_{a,1}c_{0,1}q_{e,i}} \,, \quad K_{d,i}=\frac{k_{d,i}q_{e,1}}{k_{a,1}c_{0,1}} \,, \quad 
\beta=\frac{q_{e,i}c_{0,i}}{q_{e,1}c_{0,1}}
\,.
\end{align}
The mass balance for the carrier fluid has been reduced to \eqref{cNmultireduxND}, which in turn leads to the expressions for pressure and velocity (\ref{pmultireduxND}, \ref{umultireduxND}), details are given in \ref{App:variableu}: Reduced model.
The initial and boundary conditions are
\begin{align}
    &\hat c_i(\hat x,0)= \hat q_i(\hat x,0) = 0\,, \label{ICreduxND}\\
    &\hat c_i(0,\hat t) = H(\hat t)-H(\hat t-\hat t_1) \quad \text{for}\quad i=1,...,N-1 \,, \label{BCx0reduxND}\\
    &\hat c_N(0)=1\,,\label{BCcN0reduxND}
\end{align}
the last condition \eqref{BCcN0reduxND} stems from $\hat{u}(0)=1$.

As before the mass balance and kinetic equations are uncoupled for each component and so only the solution for a single pair is required. The velocity and pressure throughout the column only depend on $x$, not on time nor any analyte concentration which simplifies  the solution.

\section{Solution methods}

We note that Laplace transforms have been applied previously in the literature. Guiochon and  Lin \cite{Guiochon} discuss a system of the form (\ref{BaseC}, \ref{BaseQ}) with a pulse inlet condition. Noting that the inverse transform is complicated  they reduce the inlet condition to a delta function. They go on to discuss the case with diffusion but are forced to apply a constant  concentration at the inlet. Their solution is taken from the work of Lapidus and Amundson \cite{Lapi52} which is valid for a semi-infinite column. In all cases the velocity remains constant throughout the column and only a single component is analysed.

\subsection{Laplace transform solution}

We now consider the system defined by equations \eqref{ArbEqs} subject to
\begin{align}
    & \hat c_i(x,0)=0, \quad \hat q_i(x,0)=0,& \forall x>0, \quad \forall i \label{eq:initial}\\
    & \hat c_i(0,t)=H(\hat t)-H(\hat t-\hat t_1), & \forall t>0, \quad \forall i. \label{eq: boundary condition}
\end{align}

\subsubsection{Solution for the constant velocity model}

Taking the Laplace transform of \eqref{ArbEqs} where  $\mathcal{L} \{c(\hat x, \hat t)\} = \Tilde{c}(\hat x, s)$  and applying the initial  conditions gives,
\begin{align}
    &s \text{Da}\Tilde{c}_i(\hat x,s)+\pad{\Tilde{c}_i}{\hat x}+s \beta_i\Tilde{q}_i(\hat x,s)=0,  \label{eq: laplace 1.1}\\
    &s \Tilde{q}_i(\hat x,s)=K_{a,i} \Tilde{c}_i -K_{d,i} \Tilde{q}_i \, ,  \label{eq: laplace 1.2}
\end{align}
subject to the boundary condition
\begin{equation}\label{laplaceBC}
    \Tilde{c}_i(0,s)=\frac{1-\exp \left(-\hat t_1s\right)}{s}.
\end{equation}

Equation \eqref{eq: laplace 1.2} provides the relation,
\begin{equation}\label{laplaceqi}
    \Tilde{q}_i(\hat x,s)=\frac{K_{a,i}}{s+K_{d,i}}\Tilde{c}_i(\hat x,s) \, .
\end{equation}
Substituting this into \eqref{eq: laplace 1.1} gives a first order differential equation for $\Tilde{c}$,
\begin{equation}
    \pad{\Tilde{c}_i}{\hat x}+s\left(\text{Da}+\frac{\beta_i K_{a,i}}{s+K_{d,i}} \right) \Tilde{c}_i =0.
\end{equation}
Integrating and applying the boundary condition determines
\begin{align}
    \Tilde{c}_i &(\hat x,s)= \label{eq: laplace sol}\\ 
    & \frac{\exp\left(-\text{Da}\hat xs\right)-\exp \left(-(\hat t_1+\text{Da}\hat x)s\right)}{s}\exp \left(\frac{\beta_iK_{a,i}K_{d,i}\hat x}{s+\Di}-\beta_iK_{a,i}\hat x\right). \nonumber 
\end{align}

The inverse transform, to return to the $t$ domain, requires use of the convolution theorem and the following results
\begin{align}
    \mathcal{L}^{-1} \left\{\exp \left(\frac{\beta_i\Aii \Di \hat x}{s+\Di} \right) \right\}&=\\
    \exp  &(-\Di \hat t) 
    \left[\sqrt{\frac{\beta_i\Aii \Di \hat x}{\hat t}} I_1(2\sqrt{\beta_i\Aii \Di \hat x \hat t})+\delta (\hat t) \right], \nonumber\end{align}
\begin{align}
    \mathcal{L}^{-1}\left\{ \frac{\exp\left(-\text{Da}\hat xs\right)-\exp \left(-(\hat t_1+\text{Da}\hat x)s\right)}{s}\right\} =  & H(\hat t-\text{Da}\hat x)-H(\hat t-\text{Da}\hat x-\hat t_1). 
\end{align}
Equation \eqref{eq: laplace sol} may then be transformed to
\begin{align}
    \hat c_i (\hat x,\hat t)& =\exp (-\beta_i\Aii \hat x) \int_0^{\hat t} \{H(\hat t-\mathcal{T}-\text{Da}\hat x)-H(\hat t-\mathcal{T}-\text{Da}\hat x-\hat t_1) \}
     \nonumber  \\ 
  \times  & \exp (-\Di \mathcal{T})\left[ \delta (\mathcal{T}) +\sqrt{\frac{\beta_i\Aii \Di \hat x}{\mathcal{T}}} I_1(2\sqrt{\beta_i\Aii \Di \hat x \mathcal{T}})\right]d\mathcal{T}.
\end{align}

The Heaviside functions restrict the   limits of integration resulting in
\begin{align}
\label{Laplaceconstu}
    \hat c_i(\hat x,t)=\int_{\max \{0, \hat t-\text{Da}\hat x-\hat t_1 \}}^{\max \{0,\hat t-\text{Da}\hat x\}}  &
    e^{-\beta_i\Aii \hat x-\Di \mathcal{T}} \times\\  & \left[ \delta (\mathcal{T}) +\sqrt{\frac{\beta_i\Aii \Di \hat x}{\mathcal{T}}} I_1(2\sqrt{\beta_i\Aii \Di \hat x \mathcal{T}})\right]d\mathcal{T}\, , \nonumber
\end{align}
where $I_1(z)$ is a modified Bessel's function of the first kind.

The problem of solving a coupled partial/ordinary differential equation system has therefore been reduced to a single numerical integration. This is easily achieved numerically.

\subsubsection{Solution for the variable velocity model}

In this case we take the Laplace transform of \eqref{cmultireduxND} and \eqref{qmultireduxND} and apply the initial conditions to find
\begin{align}
    &s \text{Da}\Tilde{c}_i(\hat x,s)+\Tilde{c}_i\pad{\hat u}{\hat x}+\hat u\pad{\Tilde{c}_i}{\hat x}+s \beta_i\Tilde{q}_i(\hat x,s)=0,  \label{eq: laplace 1.1u}\\
    &s \Tilde{q}_i(\hat x,s)=K_{a,i} \Tilde{c}_i -K_{d,i} \Tilde{q}_i. \label{eq: laplace 1.2u}
\end{align}
Equation \eqref{eq: laplace 1.2u} relates $\Tilde q$ to $\Tilde c$ which we then substitute into  \eqref{eq: laplace 1.1u} 
\begin{equation}\label{laplacedcdx}
    \pad{\Tilde{c}_i}{\hat x}+\left[\frac{1}{\hat u}\pad{\hat u}{\hat x}+\frac{s}{\hat u}\left(\text{Da}+\frac{\beta_i K_{a,i}}{s+K_{d,i}} \right)\right] \Tilde{c}_i =0\, ,
\end{equation}
where $\hat u$ is determined by 
equations (\ref{umultireduxND}, \ref{pmultireduxND}). The solution is
\begin{equation}
    \Tilde{c}_i(s,\hat x)=\frac{B(s)}{\hat u} \exp\left(\frac{s}{3\hat u_{\hat x}}\left(\text{Da}+\frac{\beta_iK_{a,i}}{s+K_{d,i}} \right)\right),
\end{equation}
where 
\begin{align}\label{dudx}
    \hat u_{\hat x}&=\pad{\hat u}{\hat x}=\frac{1-\hat p_L^2}{2\hat L\hat p^3}\, ,
\\
    B(s)&=\frac{1-\exp\left(-\hat t_1s\right)}{s}\exp\left(-\frac{s}{3\hat u_{\hat x,0}}\left(\text{Da}+\frac{\beta_iK_{a,i}}{s+K_{d,i}} \right)\right),
\end{align}
with $\hat p(\hat x)$ given by  equation \eqref{pmultireduxND} and $\hat u_{\hat x,0}=(1-\hat p_L^2)/(2\hat L)$ is the derivative of $\hat u$ evaluated at $\hat x=0$.
We thus arrive at the solution,
\begin{equation}
    \Tilde{c}_i(\hat x,s)=\frac{\exp\left(-\text{Da}\hat{\omega} s\right)-\exp \left(-(\hat t_1+\text{Da}\hat{\omega})s\right)}{\hat us}\exp \left(\frac{\beta_iK_{a,i}K_{d,i}\hat{\omega}}{s+\Di}-\beta_iK_{a,i}\hat{\omega}\right), \label{eq: laplace solu}
\end{equation}
where
\begin{equation}
    \hat{\omega}(\hat x)=\frac{1}{3}\left(\frac{1}{\hat u_{\hat x,0}}-\frac{1}{\hat u_{\hat x}}\right)=\frac{2\hat L\left(1-\hat p^3\right)}{3\left(1-\hat p_L^2\right)}\, . \label{eq: laplace omega}
\end{equation}

Again the back transform requires use of the convolution theorem
with the following results
\begin{align}
    &\mathcal{L}^{-1} \left\{\exp \left(\frac{\beta_i\Aii \Di \hat{\omega}}{s+\Di} \right) \right\}=\exp (-\Di \hat t)\left[\sqrt{\frac{\beta_i\Aii \Di \hat{\omega}}{\hat t}} I_1(2\sqrt{\beta_i\Aii \Di \hat{\omega} \hat t})+\delta (\hat t) \right],\\
    &\mathcal{L}^{-1}\left\{ \frac{\exp\left(-\text{Da}\hat{\omega} s\right)-\exp \left(-(\hat t_1+\text{Da}\hat{\omega})s\right)}{s}\right\}=H(\hat t-\text{Da}\hat{\omega})-H(\hat t-\text{Da}\hat{\omega}-\hat t_1).
\end{align}
This leads to
\begin{align}\label{Laplacesolu}
    \hat c_i(\hat x,\hat t)=\int_{\max \{0, \hat t-\text{Da}\hat{\omega}-\hat t_1 \}}^{\max \{0,\hat t-\text{Da}\hat{\omega}\}}  &
    \frac{e^{-\beta_i\Aii \hat{\omega}-\Di \mathcal{T}}}{\hat u} \times \\ & \left[ \delta (\mathcal{T}) +\sqrt{\frac{\beta_i\Aii \Di \hat{\omega}}{\mathcal{T}}} I_1(2\sqrt{\beta_i\Aii \Di \hat{\omega} \mathcal{T}})\right]d\mathcal{T} \, .\nonumber
\end{align}
In the limit of small pressure drop the pressure may be expressed as  {$\hat p \ra 1-(1-\hat p_L^2)\hat x/(2\hat L)$ and $\hat{\omega} \ra 2\hat L(1-\hat p)/(1-\hat p^2_L)$, so $\hat{\omega}  \ra \hat x$}. Equation \eqref{Laplacesolu} then reduces to \eqref{Laplaceconstu} and the constant velocity model is retrieved.

The Bessel function $I_1$ increases rapidly with increasing argument which can cause problems with the numerical integration. To avoid this we note that
\begin{align}
    I_1(z) = \frac{z}{\pi}\int_0^{\pi} \exp(z \cos \xi) \sin^2 \xi \, d\xi \, ,
\end{align}
see \cite[Eq. 9.6.18]{Abram}.
Taking the $\delta$ function outside of the integral and replacing the Bessel function we obtain the following expression   which avoids the large function values
\begin{align}\label{Laplacesolu2}
    &\hat c_i(\hat x, \hat t)= \frac{\exp\left(-\beta_i K_{ai}\hat{\omega}\right)}{\hat u}\left(H\left(\max \{0, \hat t-\text{Da}\hat{\omega}-\hat t_1 \}\right)-H\left(\max \{0, \hat t-\text{Da}\hat{\omega}\}\right)\right)+ \nonumber\\
    &\frac{2\beta_i K_{ai} K_{di}\hat{\omega}}{\pi\hat u}\int_{\max \{0, \hat t-\text{Da}\hat{\omega}-\hat t_1 \}}^{\max \{0,\hat t-\text{Da}\hat{\omega}\}} \int_0^\pi
    e^{\left(-\beta_i\Aii \hat{\omega}-\Di \mathcal{T}+2\sqrt{\beta_i\Aii \Di \hat{\omega} \mathcal{T}}\cos\xi\right)} \sin^2\xi \, d\xi d\mathcal{T} \, .
\end{align}
The solution of the PDE/ODE system is now reduced to a single integration if using 
\eqref{Laplacesolu} or a double integration if using the more stable version \eqref{Laplacesolu2}.

\subsection{Numerical solution}\label{numerics}

The fixed velocity model is specified by equations \eqref{ArbEqs}, for the concentration and amount adsorbed, subject to the boundary and initial conditions \eqref{eq:initial}-\eqref{eq: boundary condition}.  {The variable velocity model is specified by equations \eqref{umultireduxDIM}--\eqref{qmultireduxDIM}, subject to boundary and initial conditions \eqref{advdiffICu}--\eqref{advdiffBCxLu}. In order to simplify the numerical solution of the variable velocity model, we note that the accumulation term Da$\ll$1 while, if the quantity of other components is much less than 
that of the carrier fluid then the diffusion of carrier fluid is negligible (compared to advection). Thus, we use the analytical expressions for pressure and velocity in equations \eqref{pmultireduxND} and \eqref{umultireduxND}, so only the concentration and amount adsorbed of the test compounds must be determined numerically. Consequently, the same scheme can be applied to both fixed and variable velocity models.} Here we employ the method of lines with finite differences for spatial discretisation.
Further, since the analytes do not interact, it is sufficient to solve for a single pair of equations, the solutions will then differ once re-dimensionalised. 
Therefore we drop the subscript $i$ notation for this section.

The domain of the problem is discretised into $N_x+1$ spatial nodes ($\{\hat x_j\}_{j=0}^{N_x}$) and $N_t+1$ temporal nodes ($\{\hat t_i\}_{i=0}^{N_t}$). Let $\hat c_j^i=\hat c(\hat x_j,\hat t_i)$, $\hat q_j^i=\hat q(\hat x_j,\hat t_i)$.
Then from the initial conditions we obtain,
\begin{equation}
    \hat c_j^0=0, \quad  \quad \hat q_j^0=0, \qquad \forall j \neq 0,
\end{equation}
and from the boundary conditions we obtain,
\begin{align}
    \hat c_0^i&=\frac{2\Delta\hat x\left(H(\hat t_i)-H(\hat t_i-\hat t_1)\right)+\text{Pe}^{-1}\left(4\hat c_1^i-c_2^i\right)}{2\Delta\hat x\hat u + 3\text{Pe}^{-1}}, \qquad \forall i\\
    \hat c_{N_x}^i&=\frac{1}{3}\left(4\hat c_{N_x-1}^i-\hat c_{N_x-2}^i\right), \qquad \forall i\,.
\end{align}
where the first-order spatial derivatives at the inlet and the outlet have been approximated as
\begin{align}
    \pad{\hat c}{\hat x}(\hat t_i,\hat x_0)&=\frac{-3\hat c_0^{i}+4\hat c_1^{i}-\hat c_{2}^i}{2\Delta \hat x}\,, \\
    \pad{\hat c}{\hat x}(\hat t_i,\hat x_{N_x})&=\frac{3\hat c_{N_x}^{i}-4\hat c_{N_x-1}^{i}+\hat c_{N_x-2}^{i}}{2\Delta \hat x}\,.
\end{align}

 For the spatial coordinates $j=1,...,N_x-1$, we approximate the time derivatives with forward differences and the first-order and second-order spatial derivatives with a central difference,
\begin{align}
    &\pad{\hat c}{\hat t}(\hat t_i,\hat x_j)=\frac{\hat c_j^{i+1}-\hat c_j^i}{\Delta \hat t}, &
    &\pad{\hat q}{\hat t}(\hat t_i,\hat x_j)=\frac{\hat q_j^{i+1}-\hat q_j^i}{\Delta \hat t},\\
    &\pad{\hat c}{\hat x}(\hat t_i,\hat x_j)=\frac{\hat c_{j+1}^{i}-\hat c_{j-1}^i}{2\Delta \hat x}, &
    &\padd{\hat c}{\hat x}(\hat t_i,\hat x_j)=\frac{\hat c_{j+1}^{i}-2\hat c_{j}^{i}+\hat c_{j-1}^i}{\Delta \hat x^2},
\end{align}

This provides the scheme,
\begin{align}
    R_j^i= &\Aii \hat c_j^i-\Di \hat q_j^i, \qquad 
    \hat q_j^{i+1}=\hat q_j^i+\Delta \hat t R_j^i,\\
    \hat c_j^{i+1}=&\hat c_j^i\frac{\Delta \hat t}{\text{Da}} \Bigg(\hat{u}_{\hat x,j}\hat c_j^i+\left(\hat{u}_j+\text{Pe}^{-1}\mathcal{D}_{\hat x,j}\right)\frac{\hat c_{j+1}^i-\hat c_{j-1}^i}{2\Delta \hat x}+\\
    &\text{Pe}^{-1}\mathcal{D}_j\frac{\hat c_{j+1}^{i}-2\hat c_{j}^{i}+\hat c_{j-1}^i}{\Delta \hat x^2}+\beta R^i_j \Bigg), \label{cNumericsu}
\end{align}
where $\hat{u}_j$ is the value of $\hat{u}$ at $\hat x_j$, and $\hat{u}_{\hat x,j}$ is the value of the derivative $\hat{u}_{\hat{x}}=\partial \hat{u}/\partial \hat{x}$ (which can be obtained analytically) at $\hat x_j$. Since the diffusion coefficient depends on pressure, the parameter $\mathcal{D}(\hat x)=D(\hat x)/D_0$ is defined as the ratio between the diffusion coefficient at  pressure $\hat p(\hat x)$ and the diffusion coefficient at the inlet pressure ($D_0$). Since the expression for $\hat p(\hat x)$ is analytical \eqref{pmultireduxND}, we can obtain an analytical expression for $D(x)$ using the equation reported in the Appendix B of the work by Rodríguez \textit{et al.} \cite{cuevas2021numerical} (which they attribute to Ferziger and Kaper \cite{ferziger1972}). The equation indicates that the diffusion coefficient is inversely proportional to pressure, so if temperature is constant,  we can write
\begin{align}
    \mathcal{D}(\hat x)=1/\hat p(\hat x). \label{Dvsp}
\end{align}
Thus, $\mathcal{D}_{\hat x,j}$ is the value of the derivative $\mathcal{D}_{\hat{x}}=\partial \mathcal{D}/\partial \hat{x}$ at $\hat x_j$.

Note that for fixed velocity we have $\hat{u}=1$ and $\mathcal{D}=1$ for any value of $\hat x$, so equation \eqref{cNumericsu} reduces to
\begin{align}
    \hat c_j^{i+1}=\hat c_j^i-\frac{\Delta \hat t}{\text{Da}} \left(\frac{\hat c_{j+1}^i-\hat c_{j-1}^i}{2\Delta \hat x}+\text{Pe}^{-1}\frac{\hat c_{j+1}^{i}-2\hat c_{j}^{i}+\hat c_{j-1}^i}{\Delta \hat x^2}+\beta R^i_j \right). \label{cNumerics}
\end{align}

\section{Results}

\subsection{Verification}

We begin by analysing the predictions of the two solution forms: the Laplace solution and the numerical solution. To ensure that the models operate within a physically realistic parameter regime, we use the parameter values from the experiments discussed in \S \ref{NumSec}. These correspond to the experiments of \cite{cuevas2021numerical, Nasreddine2016} involving five analytes where operating conditions are provided in Tables \ref{tab:Cuevasval}, \ref{tab:Cuevasparam}. Since the experiments involve a high-pressure drop along the column, we have used the variable velocity solution, as given in equation \eqref{Laplacesolu2}.

\begin{figure}[H]
  \centering
  \begin{overpic}[width=0.508\textwidth,trim={3.4cm 2.5cm 6.5cm 1cm},clip]{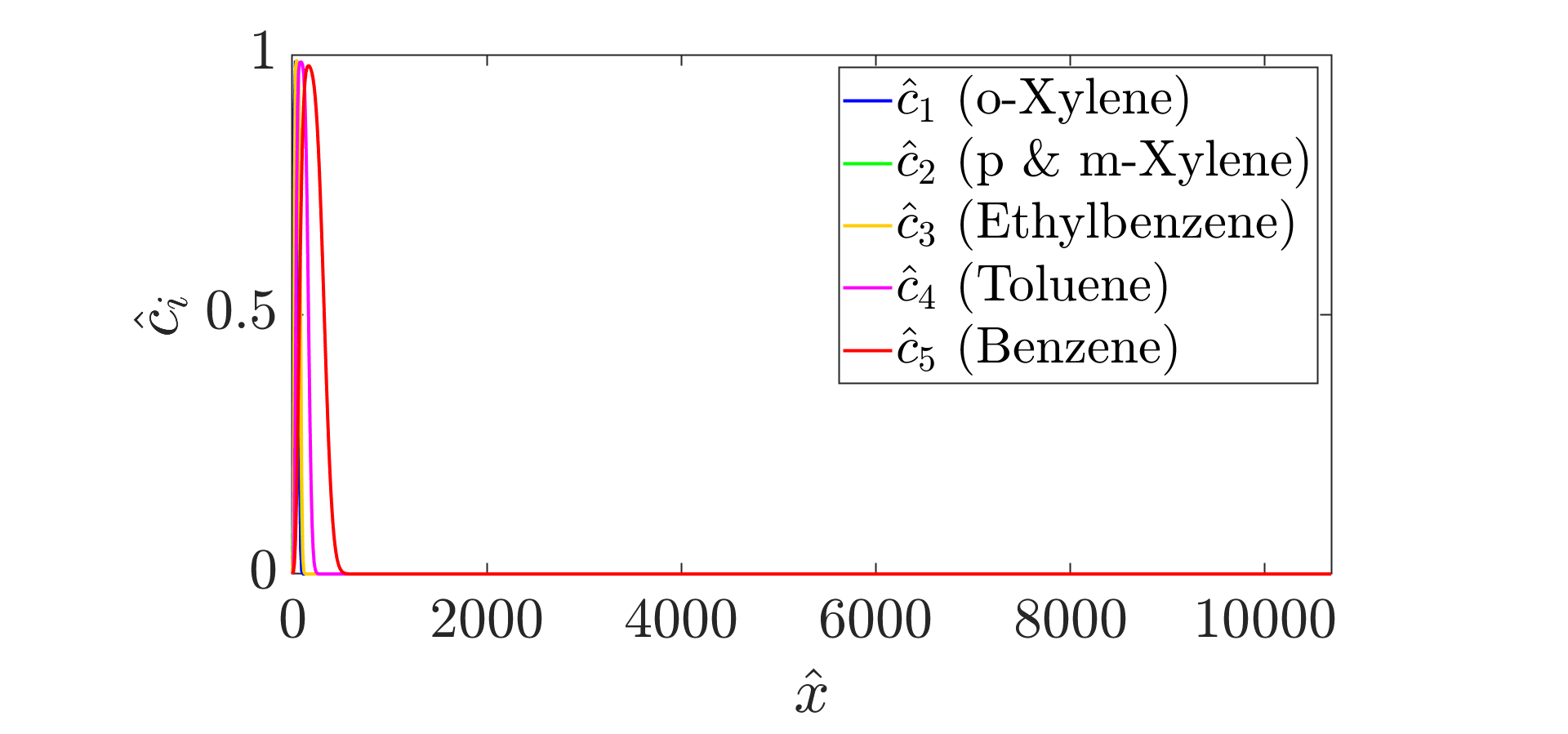}
  \put(83,9){\scriptsize $\hat{t}=70$}
  \end{overpic}\hspace{0.1cm}
  \begin{overpic}[width=0.475\textwidth,trim={6cm 2.5cm 6.5cm 1cm},clip]{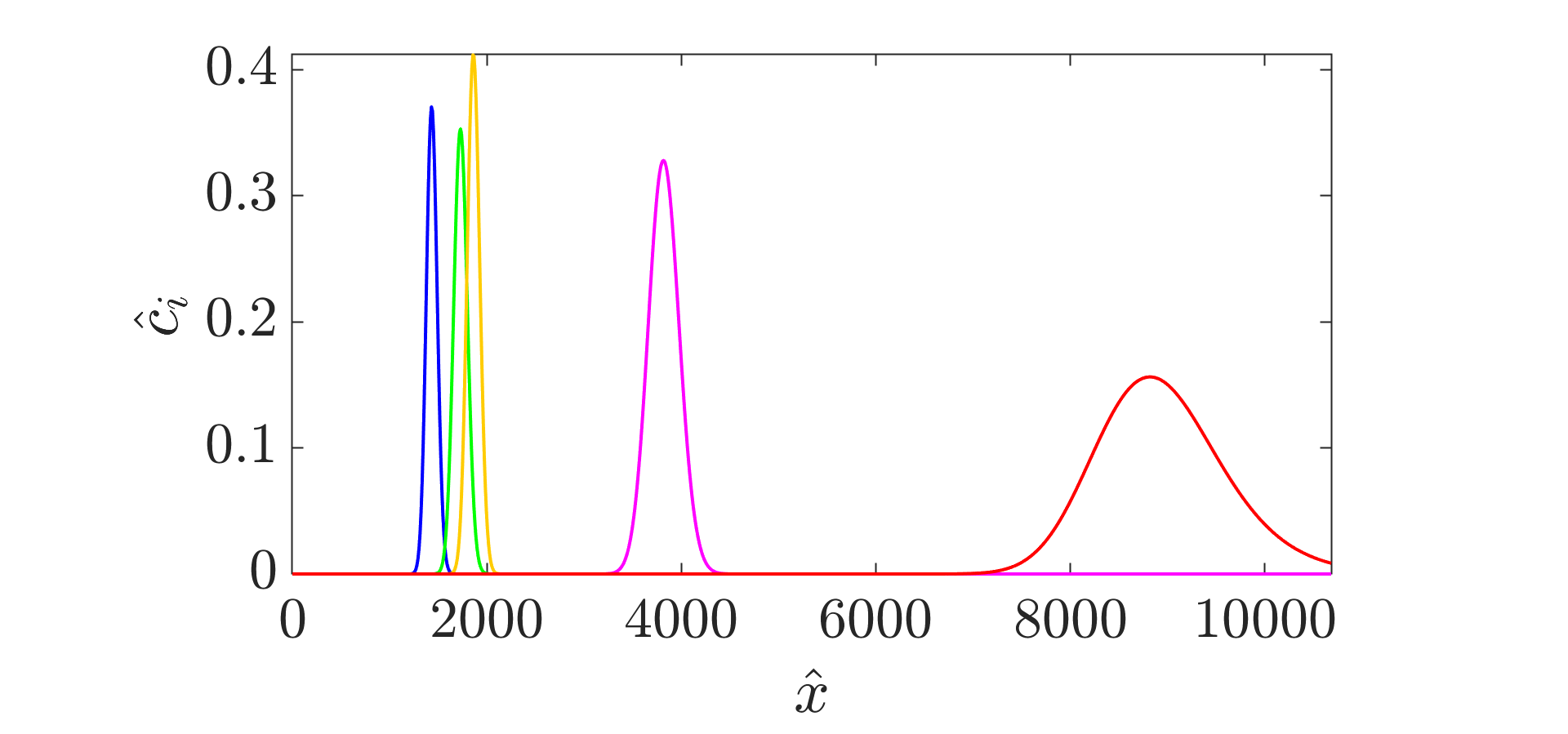}
  \put(77.7,9.5){\scriptsize $\hat{t}=1500$}
  \end{overpic}\\
  \begin{overpic}[width=0.508\textwidth,trim={3.4cm 0.5cm 6.5cm 1cm},clip]{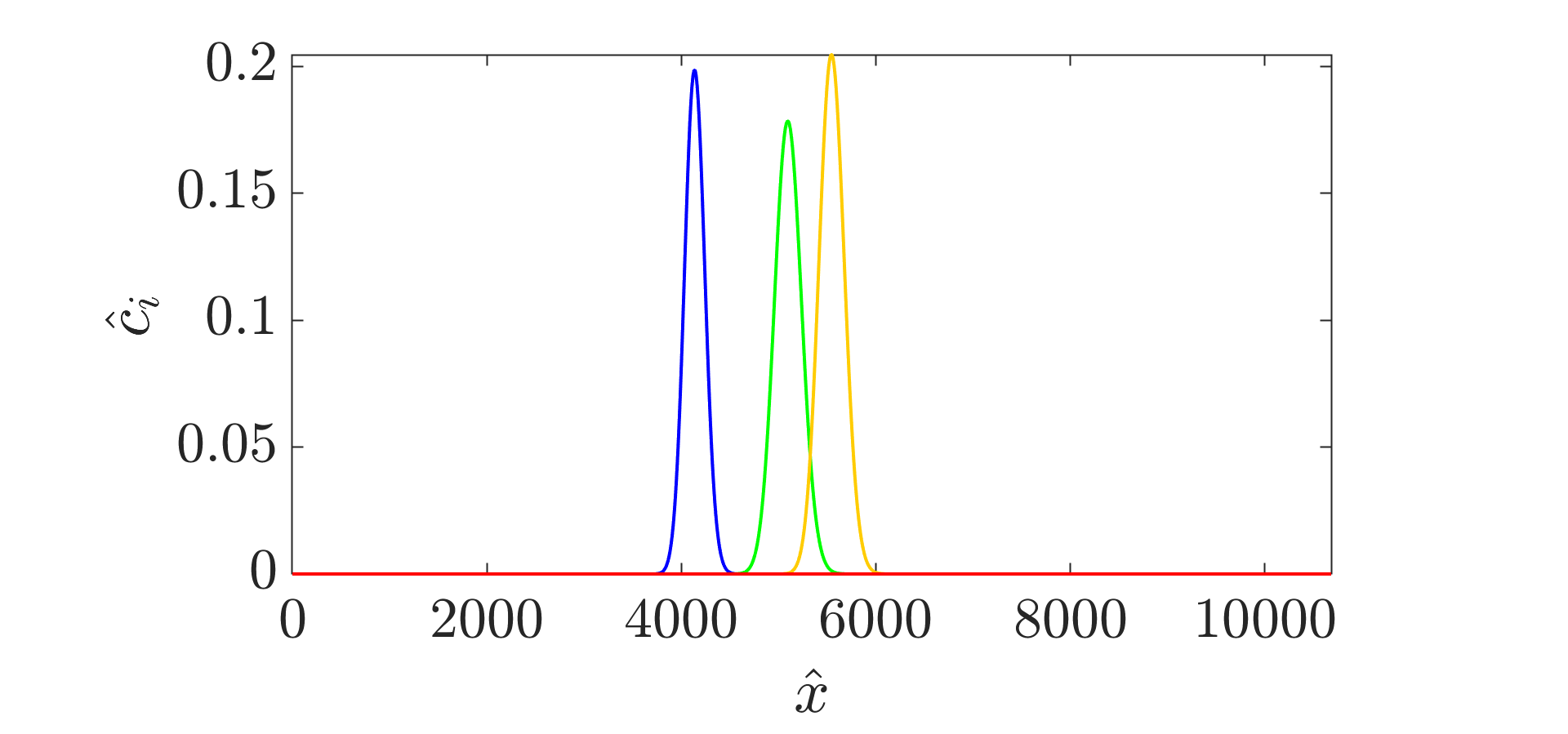}
  \put(79,14){\scriptsize $\hat{t}=4000$}
  \end{overpic}
  \begin{overpic}[width=0.481\textwidth,trim={5.5cm 0.5cm 6.5cm 1cm},clip]{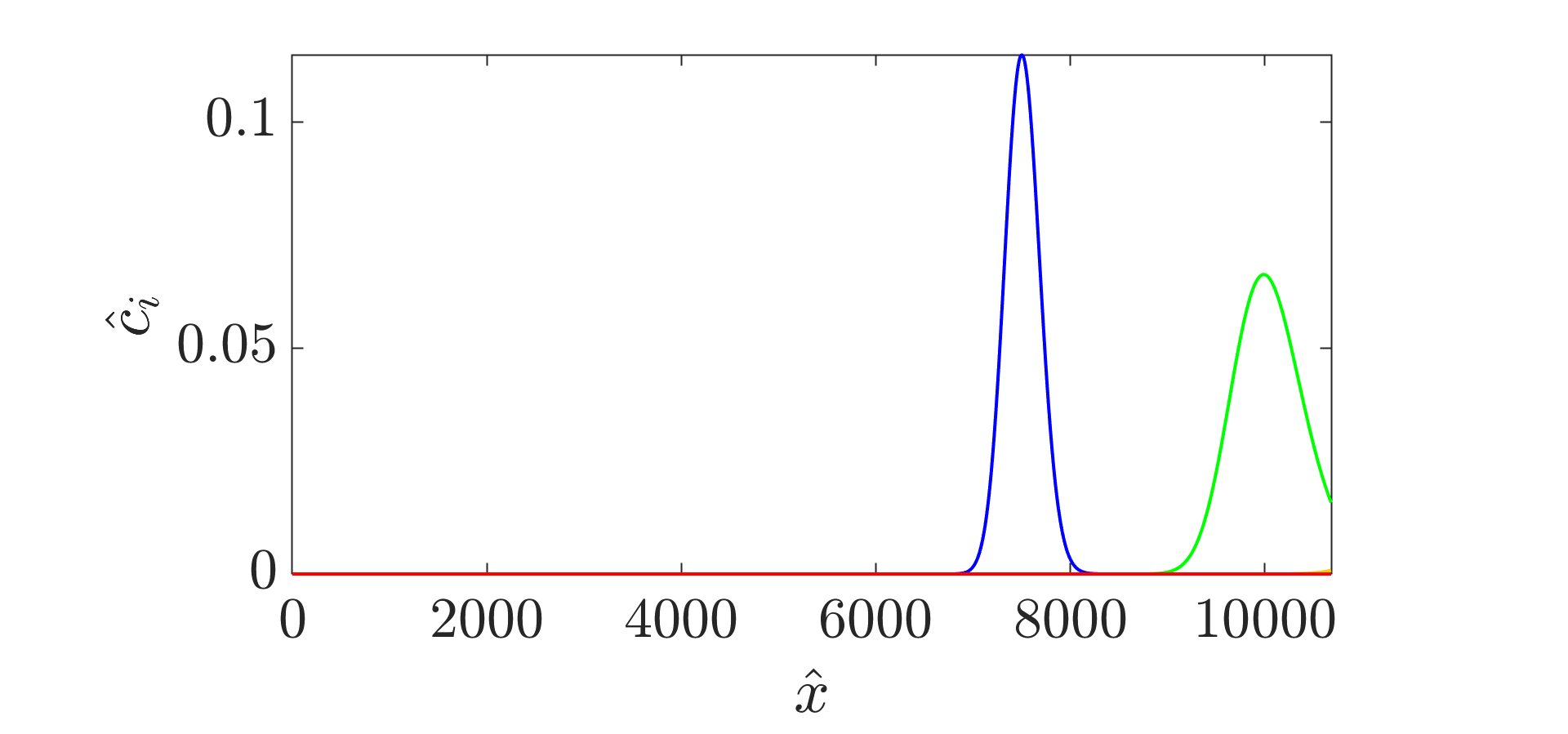}
  \put(78.5,14.5){\scriptsize $\hat{t}=6500$}
  \end{overpic}
  \caption{Simulation using Laplace solution \eqref{Laplacesolu2} for the concentration of each compound throughout the column at different times. From left to right and top to bottom: dimensionless times $\hat{t}=70$, $1500$, $4000$ and $6500$ (corresponding to dimensional $t=5$, 108, 290 and 470 s). Parameter values  are provided in Tables \ref{tab:Cuevasval} and \ref{tab:Cuevasparam}. The outlet is located at  $\hat{L}=10683$. }
  \label{fig:numericalxt}
\end{figure}

In Figure \ref{fig:numericalxt} we show the evolution of five compound concentrations along the GC column.  
At $\hat t=70$ (5s) the components are all gathered around the column inlet, however  by $\hat t = 1500$ (108s) the separation is clear: benzene travels the fastest and is close to the column outlet while xylene is the slowest. By $\hat t = 4000$ (290s) both benzene and toluene have escaped, by $\hat t = 6500$ (470s)  only o-, p-, m-xylene remains. Note that the vertical axis decreases with time, reflecting the fact that signals spread out and decrease in height as they move down the column. 

\begin{figure}[H]
  \centering
  \includegraphics[width=1\textwidth,trim={2.5cm 4.2cm 3.2cm 6.0cm},clip]{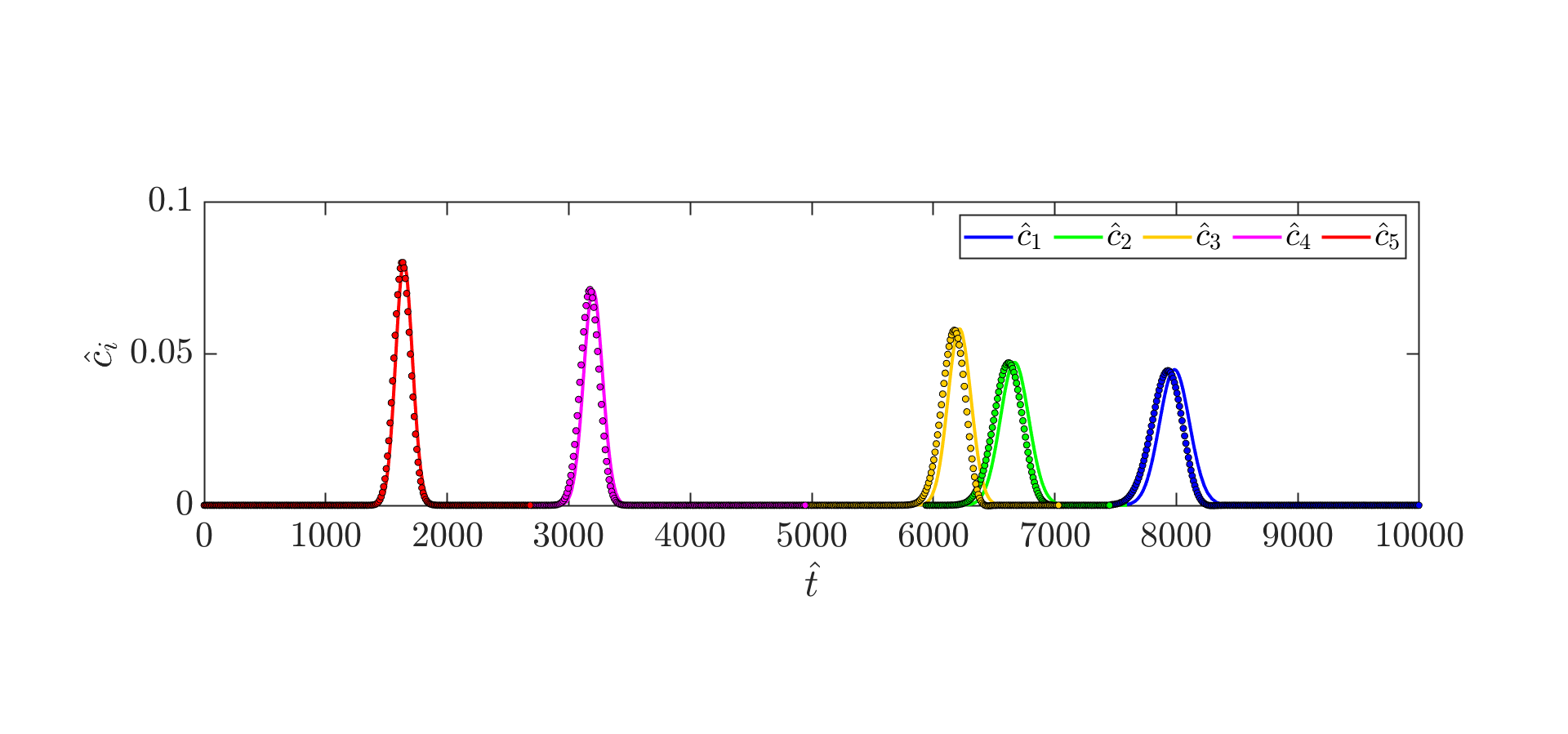}
  \includegraphics[width=1\textwidth,trim={2.5cm 4.4cm 3.2cm 6.0cm},clip]{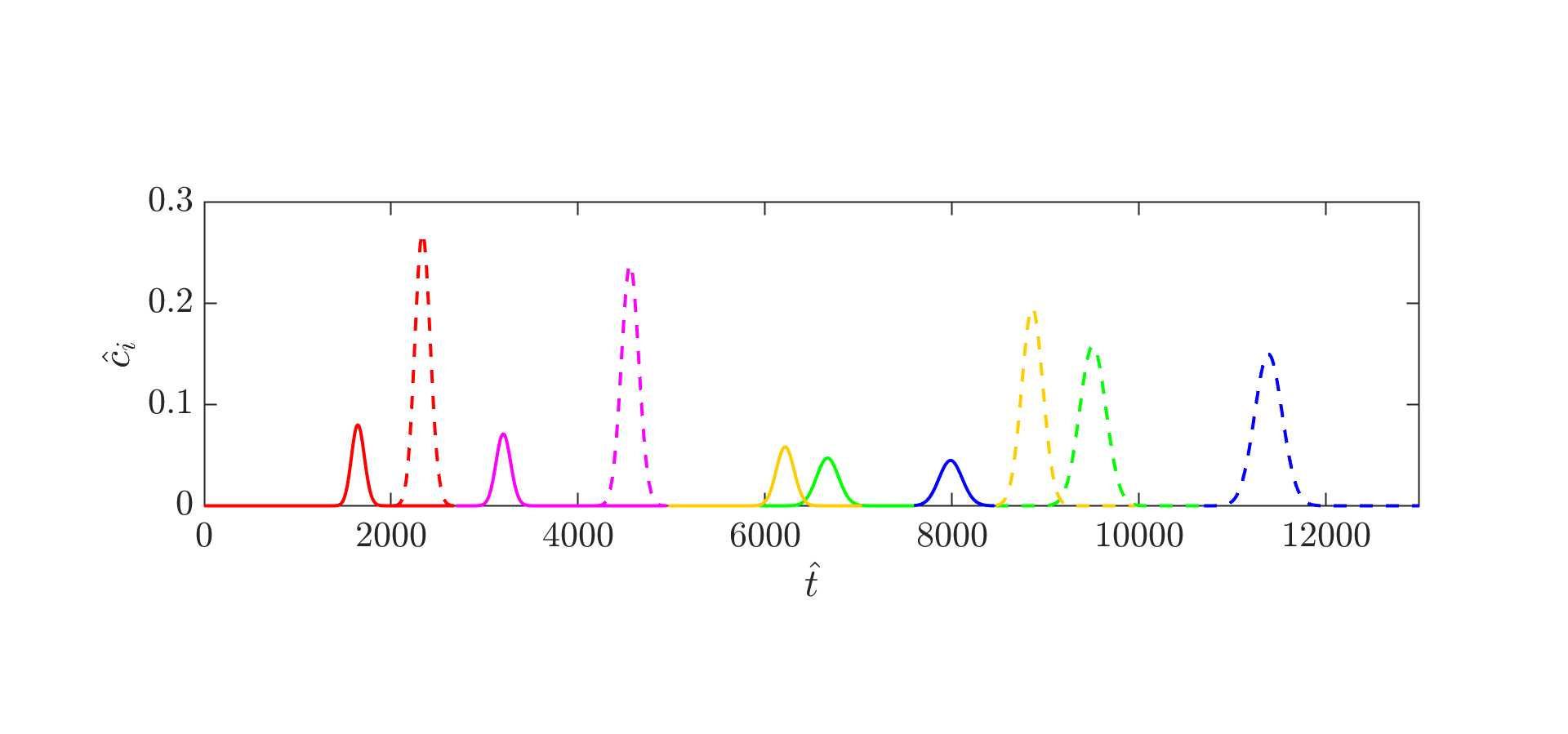}
  \caption{Comparison of the chromatograms obtained using different simulations. Top: simulation using Laplace solution \eqref{Laplacesolu2} (solid line) against numerical solution of the full PDE system in \eqref{umultireduxDIM} to \eqref{qmultireduxDIM} (circles).  Bottom: simulation using the Laplace solution of the variable velocity model \eqref{Laplacesolu2} (solid line) and the constant velocity model \eqref{Laplaceconstu} (dashed line).}
  \label{fig:numvsLaplace}
\end{figure}

In Figure \ref{fig:numvsLaplace} we compare solutions through the chromatogram (the concentration measured at the outlet). In the top figure the solid lines correspond to the Laplace solutions shown in Figure \ref{fig:numericalxt}, and the dashed lines correspond to the numerical solution of \S \ref{numerics}. As may be observed from the figure benzene escapes the column first and so shows up at the earliest time in the chromatogram while o-xylene only appears at around $\hat t \approx 7500$. The longer the residence time the more the signals spread out and so decrease in height. These two sets of results clearly demonstrate the accuracy of the Laplace solution, in comparison to the numerical one.
The numerical solution includes the effects of diffusion where 
$D_i$ obtained from Rodríguez \textit{et al}. \cite{cuevas2021numerical} (leading to Pe$_i^{-1}=0.0027-0.015$). The Laplace solution is an approximation in that it neglects diffusion, based on the observation that the inverse Peclèt number is small (indicating diffusion is negligible in comparison to advection). 
It has previously been shown that in adsorption columns, which have a higher Pe$^{-1}$ value due to dispersion, the diffusion term may be neglected without losing accuracy \cite{myers2020mass,myers2020mathematical,myers2023development}.
The close correspondence between results verifies this approximation in GC.
Consequently, henceforth we will employ only the variable velocity Laplace solution, which is significantly faster to compute.

The bottom figure shows a comparison between the constant and variable velocity Laplace solution, as given in the equations 
(\ref{Laplaceconstu}, \ref{Laplacesolu2}). 
The difference in both position and peak size is apparent.  The constant velocity model assumes that the velocity of the fluid remains constant throughout the entire column. In reality the velocity increases by a factor of 4 due to the pressure drop from 4 to 1 bar. The decrease of
pressure at the outlet also leads to lower concentrations, which can be observed through the smaller peak areas of the variable velocity model. This demonstrates that for a practical system, at least that of \cite{cuevas2021numerical, Nasreddine2016}, the constant velocity model may be highly inaccurate. However, it may be appropriate for systems with a small pressure drop or much shorter columns (for example those used in contaminant capture). 

\begin{figure}[H]
  \centering
  \includegraphics[width=0.46\textwidth]{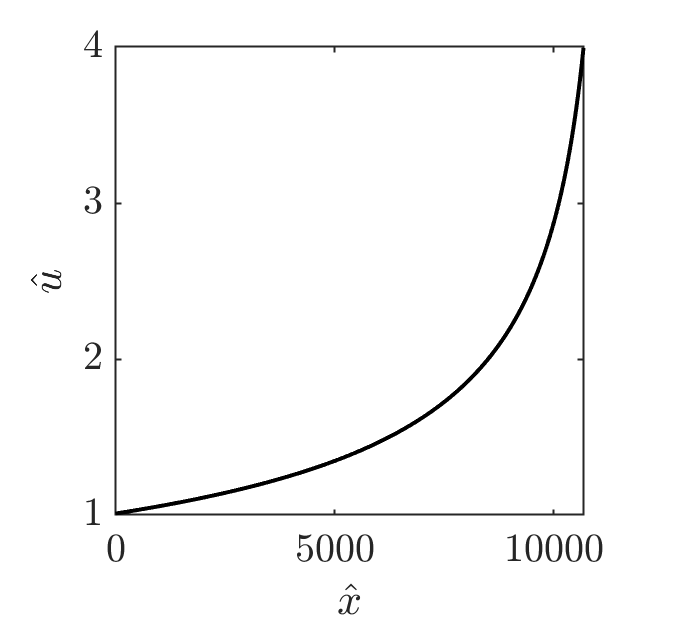} \includegraphics[width=0.46\textwidth]{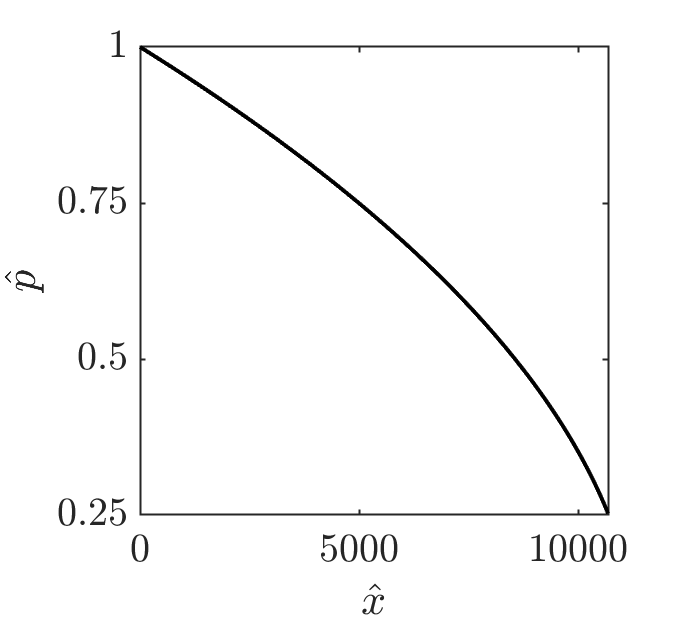}
  \caption{Evolution of velocity (left) and pressure (right) throughout the column. Results obtained using the values in Tables \ref{tab:Cuevasval} and \ref{tab:Cuevasparam}.}
  \label{fig:up}
\end{figure}

The evolution of velocity and pressure throughout the column is presented in Figure \ref{fig:up}. The large variation in velocity and pressure along the column clarifies the size of the error obtained when applying the constant velocity model to the data of \cite{cuevas2021numerical, Nasreddine2016}.

\subsection{Comparison with experiment} \label{NumSec}

Having verified the accuracy of the variable velocity Laplace solution against the numerical solution, we now verify the model against experimental data. We follow the experimental studies of  Nasreddine \textit{et al}. \cite{Nasreddine2016} and Rodríguez \textit{et al}. 
 \cite{cuevas2021numerical}. Their work involved five analytes: o-xylene (1); p-xylene and m-xylene (2); ethylbenzene (3), toluene (4); benzene (5).
The carrier fluid was nitrogen and the  inlet concentration was the same for all compounds - however since p- and m-xylene act as a single component in the GC this composite compound is assigned double the inlet concentration.  
Operating conditions are provided in Table \ref{tab:Cuevasval} for the experiment of \citealp[Fig. 6]{Nasreddine2016} (the curve labelled 20ppb). 

\begin{table}[H]
    \centering
    \caption{Operating conditions of lab-scale chromatography experiments reported by \cite{Nasreddine2016, cuevas2021numerical}.  Diffusion coefficients, at the inlet pressure, taken from \cite{cuevas2021numerical}.}
    \label{tab:Cuevasval}
    \begin{tabular}{c|c|c|c}  
        \textbf{Name} & \textbf{Symbol} & \textbf{Units} & \textbf{Value}  \\ 
        \hline
        Temperature & $T$ & K &  {353.15}  \\
        Inlet pressure & $p_0$ & Pa &  {4.01$\times$10$^5$}  \\
        Outlet pressure & $p_L$ & Pa &  {1.013$\times$10$^5$}  \\
        Column length & $L$ & m &  {20}  \\
        Inner radius & $R$ & m &  {9$\times$10$^{-5}$}  \\
        Wall thickness & $\delta$ & m &  {10$^{-6}$}  \\
        Inlet flow rate & $Q_0$ & m$^3$/s &  {10$^{-8}$}
        \\
        Inlet velocity & $u_0$ & m/s &  {0.41}
        \\
        Inlet concentration & $c_{0,i}$ & mol/m$^3$ &  2.732$\times$10$^{-6}$ (20 ppb)
        \\
        Injection time & $t_1$ & s &  {4}
        \\
        Fluid viscosity  & $\mu$ & Pa$\cdot$s &  {2.3$\times$10$^{-5}$}
        \\
        Diffusion coef.  & $D_{0,i}$ & m$^2$/s &  {2.06$\times$10$^{-6}$--2.89$\times$10$^{-6}$}
    \end{tabular}
\end{table}

Certain elements of Table \ref{tab:Cuevasval} require clarification. 
The flow rate indicated in  \cite{Nasreddine2016} is 2.5 mL/min. This was recorded at the outlet, at atmospheric pressure. In the Table we provide the value adjusted for the inlet which is at a much higher pressure,  0.625 mL/min. In \cite{LaraLbeas2019} the same research group reported that the injection time in their previous work \cite{Nasreddine2016} was 20 s. However, in Rodríguez \textit{et al}. \cite{cuevas2021numerical}  a more detailed description of the injection mechanism is provided, indicating that almost all the BTEX concentration is injected during the first 4 s. This is then the injection time taken in our calculations.

Before comparing the results with the experimental data, we need to convert the GC units. The experimental chromatogram of \cite{Nasreddine2016} has intensity units a.u. (arbitrary units). To convert these to concentration units (mol/m$^3$) we need the calibration of the gas chromatograph for each component. This requires the relation between the peak area of the chromatogram and the inlet concentration. Although Nasreddine \textit{et al}. \cite{Nasreddine2016} already report the calibration curves, we have calculated the area-concentration ratio for the experiment with the conditions from Table \ref{tab:Cuevasval}. Once this is known, we can calculate the factor between a.u. and mol/m$^3$ by assuming that the molecules injected are equal to the molecules eluted, namely
\begin{equation}\label{factorGC}
    f_ic_{0i}t_1u_0=u_L\int_0^{t_f}I_i(t)\text{d}t\,,\quad f_i=\frac{A_ip_0}{c_{0,i}t_1p_L}\,,
\end{equation}
where $f_i$ (a.u./(mol/m$^3$)) is the conversion factor for each component $i$, $u_L$ the velocity at the outlet, $t_f$ the final time of experiment (s), $I_i$ the intensity of component $i$ in the chromatogram (a.u.), and $A_i$ the area of the peak of component $i$ (a.u.$\cdot$s). Note that we have used the relation $u_0/u_L=p_L/p_0$, which stems from equation \eqref{umultireduxND}. The calibration in a.u.$\cdot$s/ppb and the conversion factor $f$ are presented in Table \ref{tab:CuevasCal}.

\begin{table}[H]
    \centering
    \caption{Calibration value (peak area/inlet concentration) and conversion factor $f$ for the chromatography experiments reported by Nasreddine \textit{et al}. \cite{Nasreddine2016} with the conditions in Table \ref{tab:Cuevasval}. The analytes with their component number are o-xylene (1), p-xylene \& m-xylene (2), ethylbenzene (3), toluene (4), benzene (5).}
    \label{tab:CuevasCal}
    \begin{tabular}{c|c|c c c c c|}  
        \multirow{2}{*}{\textbf{Parameter}}  & \multirow{2}{*}{\textbf{Units}} & \multicolumn{5}{ c |}{\textbf{Compound}} \\ 
        \cline{3-7}
        & & 1 & 2 & 3 & 4 & 5 \\
        \hline
       Calibration & a.u.$\cdot$s/ppb & 25.82 & 35.28 & 28.10 & 46.36 & 57.97  \\
        $f$ ($\times$10$^{8}$) & a.u./(mol/m$^3$) & 1.8708 & 2.5557 & 2.0356 & 3.3584 & 4.1997
    \end{tabular}
\end{table}

The calibration and the conversion factor is the same for p and m-xylene (component 2), since they contribute together to the area of the peak. Note that the values of the calibrations are very similar to the slope of the calibration curves reported by Nasreddine \textit{et al}. \cite{Nasreddine2016}.
Once the conversion factors are determined, we can fit the chromatogram obtained with the model to the experimental data.

Rodríguez \textit{et al}. \cite{cuevas2021numerical} develop a numerical solution for GC which was verified using the data of \cite{Nasreddine2016}. Their mathematical model takes a similar form to the system 
(\ref{cmultireduxDIM}, \ref{umultireduxDIM}) but with the key difference that the adsorption term for each component involves a summation which shows that all compounds compete for the same adsorption sites. This term makes it impossible to uncouple the equations and so their system must be solved simultaneously for all compounds. Further, they require a value for the maximum number of available sites - this is approximated using data from similar experiments of previous authors. They note that the value obtained is ``several orders of magnitude higher than the number of
molecules that we are injecting in the column" and consequently any errors, even of ``one or two orders of magnitude it would make no difference in the overall results".
The scheme uses finite volumes to integrate the PDE system and then an iterative scheme is applied to fit each $k_{a,i}, k_{d,i}$ to the experimental data of all components.
The process of solving simultaneously the PDE system for all five components and then iterating for ten unknowns involves a high computational cost. 

In our approach we have decoupled the equations, based on the assumption there are sufficient available adsorption sites at all times and so the competition is negligible. This is clearly the case in the latter stages of the process when the components have separated. However, the findings of \cite{cuevas2021numerical}, that the number of sites is orders of magnitude greater than the number of injected molecules, vindicates this assumption and makes it clear that the competition term is negligible. Our solution, equation \eqref{Laplacesolu2}, involves a single integration to determine the concentration profile for all components (since they are scaled versions of each other).  Substituting $x_L$ then determines the concentration at the outlet, which defines the chromatogram. For a given analyte we then only need to fit two parameters to the data: there is no need to calculate the maximum number of adsorption sites. This results in a highly efficient solution method. 

\begin{figure}[H]
  \centering
  \includegraphics[width=1\textwidth]{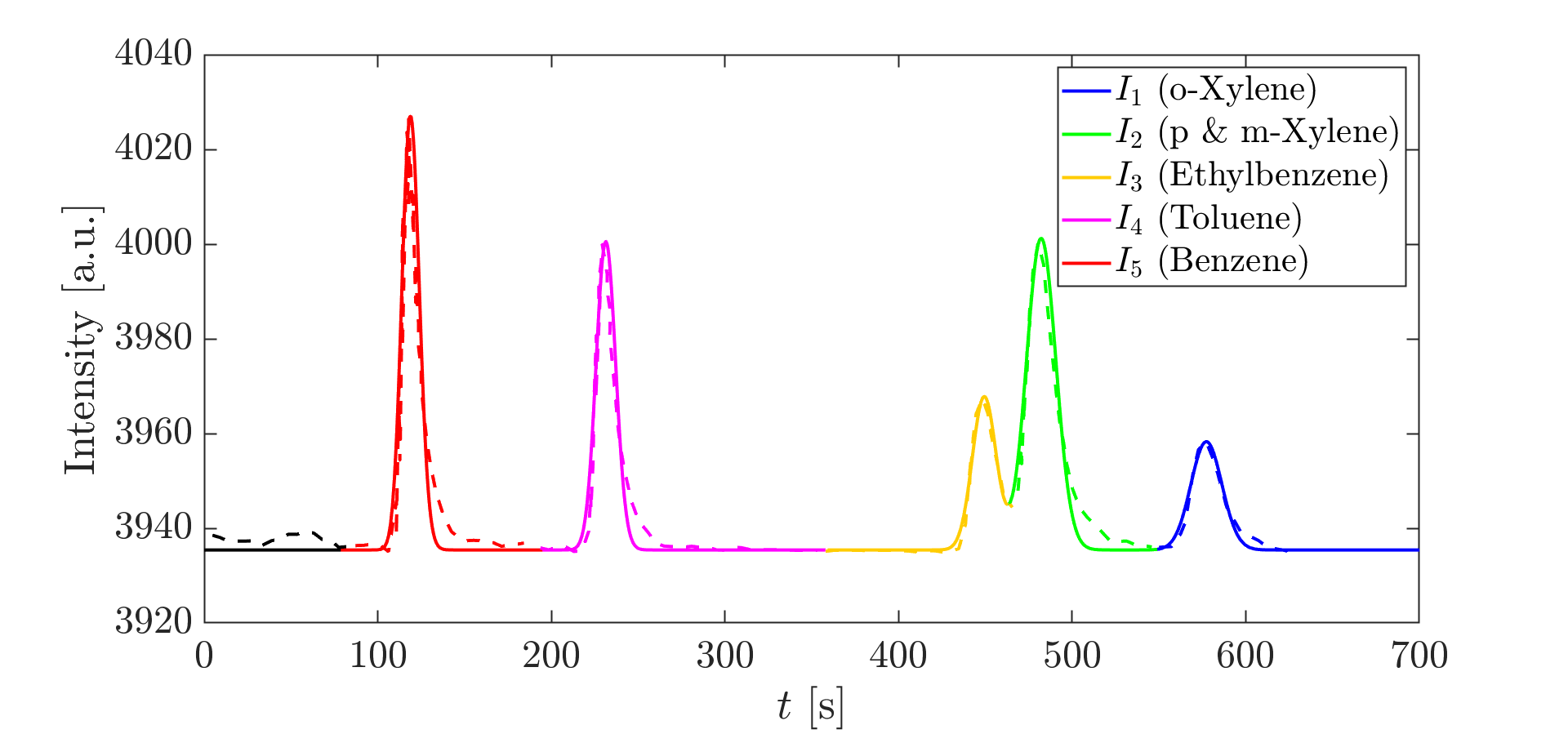}
  \caption{Fitting of the variable velocity model (solid line) to the experimental chromatograph (striped line) reported by Nasreddine \textit{et al}. \cite{Nasreddine2016}. Results obtained using the values in Tables \ref{tab:Cuevasval} and \ref{tab:Cuevasparam}.}
  \label{fig:fitting}
\end{figure}

In Figure \ref{fig:fitting} we show the result of matching the Laplace solution to the experimental data \cite{Nasreddine2016}. Note that the experimental data has a non-zero baseline. This background is approximately constant, so the mean value (3935.4 a.u.) has been calculated from the flat region between component 3 and 4 (between 300 and 400 s approximately). This value has then been added to the simulation results when fitting the experimental data. Equation \eqref{Laplacesolu2} was integrated using a global adaptive quadrature method. Then, the Matlab global optimisation function \textit{GlobalSearch} was used to fit the model to the experimental data. The local solver coupled to \textit{GlobalSearch} uses interior-point algorithm to minimise the sum of squared errors (SSE). 
The chromatogram shows an overlap between component 2 (ethylbenzene) and 3 (toluene). However, the overlap region is small, we neglected this small section in the data fitting.

As shown in Figure \ref{fig:fitting}, the agreement between the analytical solution, equation \eqref{Laplacesolu2}, and the experimental data is very good. 
The fitting parameters to achieve this, $k_{a,i}, k_{d,i}$, are presented in Table \ref{tab:Cuevasparam}. Also shown are a number of related parameters, which may be subsequently calculated with the aid of values provided in Table \ref{tab:Cuevasval}.  With regard to the discussion concerning goodness of fit we note that the $R^2$ values provided in the table are all high. The main deviations between the solutions and data are due to the peak tailing. The causes of the  tailing are  discussed in  \cite{cuevas2021numerical}, where it is suggested that the culprit is the flow path disruption at the inlet of the column. This leads to some molecules   entering the column after the initial four seconds.

\begin{table}[H]
    \centering
    \caption{Physical parameters obtained from fitting the model to the experimental data by Nasreddine \textit{et al}. \cite{Nasreddine2016}. The only fitting parameters are $k_{a,i}$ and $k_{d,i}$. The rest may be calculated from  $k_{a,i}$, $k_{d,i}$ and parameters  provided in Table \ref{tab:Cuevasval}. The analytes with their component number are o-xylene (1), p-xylene \& m-xylene (2), ethylbenzene (3), toluene (4), benzene (5).}
    \label{tab:Cuevasparam}
    \begin{tabular}{c|c|c c c c c|}  
        \multirow{2}{*}{\textbf{Parameter}}  & \multirow{2}{*}{\textbf{Units}} & \multicolumn{5}{ c |}{\textbf{Compound}} \\ 
        \cline{3-7}
        & & 1 & 2 & 3 & 4 & 5 \\
        \hline
        $k_{a,i}$ ($\times$10$^{4}$) & $s^{-1}$ & 1.0168 & 0.6483 & 0.9203 & 0.2953 & 0.0579  \\
        $k_{d,i}$ & $s^{-1}$ & 14.291 & 11.057 & 16.940 & 11.502 & 5.299 \\
        $\mathcal{K}_i$ & - & 711.55 & 586.33 & 543.31 & 256.73 & 109.28 \\
        $q_{e,i}$ ($\times$10$^{-3}$)& mol/m$^3$ & 1.9436 & 3.2032 & 1.4841 & 0.7013 & 0.2985 \\
        $\beta_i$ & - & 1.0000 & 0.8242 & 0.7635 & 0.3608 & 0.1530 \\
        Pe$_i^{-1}$ ($\times$10$^{-3}$) & - & 2.6926 & 2.6926 & 3.1069 & 3.3968 & 3.7696 \\
        Da & - & \multicolumn{5}{ c |}{0.0633} \\
        $\mathcal{L}$ ($\times$10$^{-3}$) & m & \multicolumn{5}{ c |}{1.8722} \\
        $\tau$ & s & \multicolumn{5}{ c |}{0.0723} \\
        SSE ($\times$10$^{3}$) & a.u.$^2$ & 0.0323 & 0.7044 & 0.1232 & 1.3718 & 6.0291 \\
        R-squared & - & 0.9715 & 0.9425 & 0.9642 & 0.9402 & 0.8585 
    \end{tabular}
\end{table}


\section{Conclusions}

This works presents a novel solution to simulate and predict gas chromatography column processes. Two different scenarios have been considered: a first situation where pressure drop is negligible, so no effects on the velocity field are considered, and a second situation where pressure and velocity vary throughout the column. 

Although the system study is analogous to others previously recorded in the literature the solution technique is novel. Standard solutions use numerical techniques to integrate the PDE system and then an iterative scheme to fit adsorption and desorption coefficients to the experimental data of all components.
The process of solving simultaneously the PDE system for all  components and then iterating for all unknowns involves a high computational cost. 
In our approach  the equations are decoupled and the solution  involves a single integration which determines the concentration profile for all components (since they are scaled versions of each other).   For a given analyte we then only need to fit two parameters to the data. This results in a highly efficient solution method. 

Comparison of our solution against numerics and experimental data verified the accuracy of the new approach. A comparison between the constant and variable velocity models demonstrated the possible inaccuracy of neglecting pressure effects.

The model outputs include the adsorption coefficient, which is related to the variance of the peaks in the chromatogram, and the equilibrium constant, which is related to the retention time. These two are key in optimising the chromatography process.

Whilst the new approach is efficient and provides excellent agreement with experimental data it must be viewed as a starting point. Many chromatography processes involve heating systems, which is not considered in our isothermal model. 
The impact of higher concentrations or other types of eluents, analytes and adsorbents are also  possible future extensions. Including these aspects would further increase the value and applicability of the methods presented in this paper. 

\section*{Acknowledgements}

First of all we would like to thank Prof. David Mason for his tireless dedication toward the South African Mathematics in Industry Study Groups. Many inspiring problems have arisen in these meetings. The work described in this paper being just one of them!

T. Myers, A. Cabrera-Codony were funded by MCIN/AEI/ 10.13039/50110 0 011033/ and by “ERDF A way of making Europe”, grant number PID2020- 115023RB-I00. 
TM acknowledges the CERCA Programme of the Generalitat de Catalunya and the Spanish State Research Agency (AEI), through the Severo Ochoa and Maria de Maeztu Program for Centres and Units of Excellence in R\&D (CEX2020-001084-M). ACC acknowledges AEI for Juan de la Cierva Incorporación fellowship (IJC2019-038874-I). 

A. Valverde acknowledges support from the Margarita Salas
UPC postdoctoral grants funded by the Spanish Ministry of Universities with
European Union funds - NextGenerationEU (UNI/551/2021 UP2021-034) 

O. A. I. Noreldin acknowledges financial support from the University of Zululand.

\appendix
\section{Cross-sectional averaging for constant velocity model} \label{App:constantu}
The standard mass balance for flow down the column is
\begin{align}\label{FirstEqs}
\pad c t + \bf u \cdot \nabla c &= D \nabla^2 c,
\end{align}
where we have assumed constant diffusivity $D$ and velocity $u = \dot M/(\pi R^2 \rho)$, where $\dot M$ is the mass flux.
We define a cross-sectional average concentration in the mobile phase
\begin{align}
\pi R^2 \bar c = 2 \pi \int_0^R r c dr.
\end{align}
Integrating the concentration equation over the cross-section
\begin{align} 
2 \pi \int_0^R \left(\pad c t + u \pad c x \right) r dr = 2 \pi D \int_0^R &\left(\padd c x + \frac{1}{r}\pad{}{r} \left( r \pad c r\right)\right) r dr \nonumber\\
\Ra \qquad \qquad \pi R^2   \left(\pad{\bar c}{t}  + u \pad{\bar c}{x} \right)& =  D \left(
\pi R^2 \padd{\bar c}{x} + 2 \pi \left. r \pad c r\right|_0^R \right) \, . \label{cInt}
\end{align}
The final term of \eqref{cInt} represents the contribution of the diffusive flux. Due to symmetry it is zero at $r=0$, at $r=R$ it represents the mass flux onto  the stationary phase. Assuming mass is evenly distributed within the thin stationary phase
\begin{align}
 2 \pi \left. R D \pad c r\right|^R = - \pi \left[(R+\delta)^2 - R^2\right]  \pad{\bar q}{t} \approx - 2 \pi R \delta  \pad{\bar q}{t} 
\end{align}
where $\bar q$  is the average concentration
of the attached molecules and neglecting $\delta/R \ll 1$. Replacing this in equation \eqref{cInt} leads to
\begin{align}\label{CInt}
\pad{\bar c}{t} + u \pad{\bar c}{x}  &=  D \padd{\bar c}{x} - \frac{2\delta}{R}\pad{\bar q}{t}.
\end{align}
Defining $\alpha = 2\delta/R$ we arrive at Eq. \eqref{Eq1}.

Appropriate initial and boundary conditions reflect the fact there is no trace of analyte in the column at $t=0$ and then the injection occurs over a period $0 \le t \le t_1$, 
\begin{align} \label{BCsAppA}
    &c(x,0)= q(x,0) = 0\,, \\
    &\left(uc-D\frac{\partial c}{\partial x}\right)\Bigg\vert_{x=0^+}= u c_{0}\left(H(t)-H(t-t_1)\right) \, , 
\end{align}
where $H$ is the Heaviside function. At the outlet we apply 
\begin{align} \label{BCLAppA}
    \frac{\partial c_i}{\partial x}\Big\vert_{x=L}=0 \,,
\end{align}
where $L$ is the length of the column (m).


\section{Derivation of variable velocity model}\label{App:variableu}

\subsection{Dimensional model}

For a long thin column the flow is primarily along the axis, such that $\textbf{u}=v(r,x,t) \textbf{e}_x$, where  $\textbf{e}_x$ is the unit vector in the axial direction.  Momentum conservation of the carrier fluid is described by the Navier-Stokes equation, assuming radial symmetry and no external forces applied, this reads
\begin{align}\label{stokeseq}
    \mu\left(\frac{1}{r}\frac{\partial}{\partial r}\left(r\frac{\partial v}{\partial r}\right)+\frac{\partial^2v}{\partial x^2}\right)=\frac{\partial p}{\partial x} \, ,
\end{align}
where $\mu$ is the dynamic viscosity of the carrier fluid (Pa$\cdot$s) and $p(x,t)$ is the pressure inside the column (Pa).

The boundary conditions of equation \eqref{stokeseq} account for  no-slip at the solid surface ($r=R$) and  radial symmetry in the center of the column ($r=0$).  Equation \eqref{stokeseq} is first order in $p$, yet it must satisfy two pressure  conditions (at the inlet $p=p_0$, at the outlet $p=p_L$). Here we choose to apply  the outlet condition, the inlet condition then determines the flow rate. Hence, we apply
\begin{align}\label{stokesBC}
    v(R,x,t)=0\,, \qquad \frac{\partial v}{\partial r}\Big\vert_{r=0}=0\,, \qquad p(L,t)=p_L \,.
\end{align}

Mass conservation for analyte follows Eq. \eqref{FirstEqs} but the advection term is written as $ (\textbf{u} c)_x = (vc)_x$. 

Due to the $vc$ term we cannot immediately carry out the averaging process and must therefore first consider the reduced non-dimensional system.

\subsection{Dimensionless model}

Following the non-dimensionalisation outlined in \S \ref{sec:modelvariableu} we have the system
\begin{align}
    &\frac{1}{\hat r}\frac{\partial}{\partial \hat r}\left(\hat r\frac{\partial \hat v}{\partial \hat r}\right)+\varepsilon^2\frac{\partial^2\hat v}{\partial \hat x^2}=\frac{\partial \hat p}{\partial \hat x} \, , \label{nondimstokes}\\
    &\text{Pe}\varepsilon^2\left(\text{Da}\frac{\partial \hat c}{\partial\hat t}+\frac{\partial (\hat v \hat c)}{\partial \hat x}\right)=\varepsilon^2\frac{\partial}{\partial \hat x}\left(\mathcal{D}\frac{\partial \hat c}{\partial \hat x}\right)+\frac{1}{\hat r}\frac{\partial}{\partial \hat r}\left(\mathcal{D}\hat r\frac{\partial \hat c}{\partial \hat r}\right) \, , \label{nondimadvdiffmulti}
\end{align}
where $\text{Pe}=u_0\mathcal{L}/D_{0}$, $\text{Da}=\mathcal{L}/(u_0\tau)$, $\varepsilon=\mathcal{R}/\mathcal{L}$ and the pressure scale  $\mathcal{P}=\mu u_0 \mathcal{L}/\mathcal{R}^2$. Since the diffusion coefficient depends on pressure and temperature, we define $\mathcal{D}=D/D_{0}$ to distinguish between the diffusion coefficient at the inlet pressure with initial temperature conditions ($D_{0}$), and the dimensionless variable function $\mathcal{D}(\hat x,\hat t)$. Note that the dynamic viscosity of the fluid $\mu$ is taken as a constant regardless of the composition of the mixture. This is based on the assumption that only trace amounts of analyte are injected in the carrier fluid.
The boundary conditions of \eqref{nondimstokes} are
\begin{align}\label{nondimstokesBC}
    \hat v(R,x,t)=0\,, \qquad \frac{\partial \hat v}{\partial \hat r}\Big\vert_{\hat r=0}=0\,, \qquad \hat p(\hat L,\hat t)=\hat p_L \,.
\end{align}
The initial and boundary conditions of equation \eqref{nondimadvdiffmulti} are
\begin{align} 
    &\hat c(\hat x,0)= \hat q(\hat x,0) = 0\,, \label{nondimadvdiffIC}\\
    &\left(\hat v\hat c-\text{Pe}^{-1}\mathcal{D}\frac{\partial\hat c}{\partial\hat x}\right)\Bigg\vert_{\hat x=0^+} = H(\hat t)-H(\hat t-\hat t_1)   \,, \label{nondimadvdiffBCx0} \\
    &\frac{\partial \hat c}{\partial \hat x}\Big\vert_{\hat x=\hat L}=0 \,,\label{nondimadvdiffBCxL}\\
    &\mathcal{D}\frac{\partial \hat c}{\partial \hat r}\Big\vert_{\hat r=\hat R}= -  \varepsilon^2\frac{\partial \hat q}{\partial \hat t}\,, \qquad \frac{\partial \hat c}{\partial \hat r}\Big\vert_{\hat r=0}=0 \,.\label{nondimadvdiffBCr}
\end{align}
 Note that if temperature is constant, $\mathcal{D}(0,\hat t)=1$.

We consider now the asymptotic expansion 
$$
f=f^{(0)}+\varepsilon^2f^{(1)}+\mathcal{O}\left(\varepsilon^4\right)\,,
$$
where $f=\{\hat c,\hat q, \hat v, \hat p\}$.
The leading order in $\varepsilon^2$ of equation \eqref{nondimstokes} reads
\begin{align}
    \frac{1}{\hat r}\frac{\partial}{\partial \hat r}\left(\hat r\frac{\partial \hat v^{(0)}}{\partial \hat r}\right)=\frac{\partial \hat p^{(0)}}{\partial \hat x} \, . \label{nondimstokeslead}
\end{align}
Integrating equation \eqref{nondimstokeslead} with boundary conditions in \eqref{nondimstokesBC} gives
\begin{align}\label{nondimstokessol}
    \hat{v}^{(0)}=\frac{\hat{r}^2-\hat{R}^2}{4}\frac{\partial \hat p^{(0)}}{\partial \hat x}\,.
\end{align}
We define the dimensionless average velocity as
\begin{align}\label{averageu}
    \hat{u}=\frac{2}{\hat R^2}\int_0^{\hat{R}}\hat v\hat r\,\text{d}\hat r\, ,
\end{align}
which applied to \eqref{nondimstokessol} gives
\begin{align}\label{nondimu}
    \hat{u}^{(0)}=-\frac{\hat{R}^2}{8}\frac{\partial \hat p^{(0)}}{\partial \hat x}\,.
\end{align}

Taking the leading order in $\varepsilon^2$ of equation \eqref{nondimadvdiffmulti} we get 
\begin{align}
    \frac{1}{\hat r}\frac{\partial}{\partial \hat r}\left(\mathcal{D}\hat r\frac{\partial \hat c^{(0)}}{\partial \hat r}\right)=0 \, . \label{nondimadvdiffmultilead}
\end{align}
Solving equation \eqref{nondimadvdiffmultilead} subject to the boundary condition on $r=0$ in \eqref{nondimadvdiffBCr}, we get $\hat c^{(0)}=\hat c^{(0)}(\hat x,\hat t)$.

The first order in $\varepsilon^2$, equation \eqref{nondimadvdiffmulti} reads
\begin{align}
    \text{Da}\frac{\partial \hat c ^{(0)}}{\partial\hat t}+\frac{\partial (\hat v^{(0)} \hat c ^{(0)})}{\partial \hat x}=\text{Pe} ^{-1}\frac{\partial}{\partial \hat x}\left(\mathcal{D} \frac{\partial \hat c ^{(0)}}{\partial \hat x}\right)+\frac{\text{Pe} ^{-1}}{\hat r}\frac{\partial}{\partial \hat r}\left(\mathcal{D} \hat r\frac{\partial \hat c ^{(1)}}{\partial \hat r}\right) \, . \label{nondimadvdiffmultifirst}
\end{align}
Integrating equation \eqref{nondimadvdiffmultifirst} over the cross-section we obtain
\begin{align}
    \text{Da}\frac{\partial \hat c ^{(0)}}{\partial\hat t}+\frac{\partial (\hat u^{(0)} \hat c ^{(0)})}{\partial \hat x}=\text{Pe} ^{-1}\frac{\partial}{\partial \hat x}\left(\mathcal{D} \frac{\partial \hat c ^{(0)}}{\partial \hat x}\right)+\frac{2\text{Pe} ^{-1}\mathcal{D} }{\hat R}\frac{\partial \hat c ^{(1)}}{\partial \hat r}\Big\vert_{\hat r=\hat R} \, . \label{nondimadvdiffmultifirstaverage}
\end{align}
The averaging  affects the boundary condition at the inlet
\begin{align} 
    &\left(\hat u^{(0)}\hat c ^{(0)}-\text{Pe} ^{-1}\mathcal{D} \frac{\partial\hat c ^{(0)}}{\partial\hat x}\right)\Bigg\vert_{\hat x=0^+} = H(\hat t)-H(\hat t-\hat t_1) \,.\label{nondimadvdiffBCx0Naverage}
\end{align}
Note that the derivative of $\hat c ^{(1)}$ at the boundary $\hat r=\hat R$ in equation \eqref{nondimadvdiffmultifirstaverage} must match with the sink term to leading order. Consequently we write
\begin{align} 
    &\mathcal{D} \frac{\partial \hat c ^{(1)}}{\partial \hat r}\Big\vert_{\hat r=\hat R}= -   \frac{\partial \hat q ^{(0)}}{\partial \hat t}\,. \label{nondimadvdiffBCrfirst}
\end{align}
Note that in order to obtain the desired order of magnitude of $\varepsilon^2$ in the sink term, the radial length-scale must be defined as $\mathcal{R}=\delta q_{e}\mathcal{L}^2/(c_{0}D_{0}\tau)$. 

Now, replacing \eqref{nondimadvdiffBCrfirst} in \eqref{nondimadvdiffmultifirstaverage} we get
\begin{align}
    \text{Da}\frac{\partial \hat c^{(0)}}{\partial\hat t}+\frac{\partial (\hat u^{(0)} \hat c^{(0)})}{\partial \hat x}=\text{Pe}^{-1}\frac{\partial}{\partial \hat x}\left(\mathcal{D}\frac{\partial \hat c^{(0)}}{\partial \hat x}\right)- \frac{\partial \hat q^{(0)}}{\partial \hat t} \, , \label{nondimadvdiffmultifinal}
\end{align}
where the length-scale has been defined as $\mathcal{L}=\tau u_0 c_{0} R/(2\delta q_{e})$. Once the length-scale is defined, the radial length-scale and the pressure scale can be rewritten as $\mathcal{R}=u_0^2R^2c_{0}\tau/(4\delta q_{e}D_{0})$ and $\mathcal{P}=8\mu\delta q_{e}D_{0}^2/\left(u_0^2R^3c_{0}\tau\right)$, and the parameter $\varepsilon=u_0R/(2D_{0})$.

The pressure of the column to leading order may be defined in terms of the concentration of the carrier fluid
\begin{align}\label{idealgasEoSNDredux}
    \hat p^{(0)}(\hat x,\hat t)=\hat p_0 \hat c_N^{(0)}\,,
\end{align}
where, anticipating the extension to multiple contaminants, we denote the concentration of the carrier fluid as $c_N^{(0)}(\hat{x},\hat{t})$.

As far as the sink term is concerned, as before we assume the adsorption capacity is much greater than the concentration of analyte and write
\begin{align} \label{LangcompleteNDlead}
    \frac{\partial \hat q^{(0)}}{\partial \hat t}= \hat c^{(0)}-K_{d}\hat q^{(0)}\, , 
\end{align}
where $K_d = k_d q_e/(k_a c_0)$.

\subsection{Reduced Model} \label{AppRedMod}

When dealing with gases, the diffusion coefficient can reach values of the order of magnitude of 10$^{-5}$ m$^2$/s \cite{cuevas2021numerical}. Even higher values have been reported in packed columns where dispersion acts to increase the value of $\mathcal{D}$. In packed columns the  effect on the concentration has been demonstrated to be negligible \cite{myers2023development}. Consequently we  assume that $\text{Pe}^{-1}\ll 1$. Neglecting diffusion reduces the order of  equation  \eqref{nondimadvdiffmultifinal}. Then boundary condition at $x=L$ doesn't hold, and only \eqref{nondimadvdiffBCx0Naverage}, neglecting the $\text{Pe}^{-1}$ term, applies.

To understand the flow of the carrier fluid we write a reduced form \eqref{nondimadvdiffmultifinal} where $Da \ll \text{Pe}^{-1}\ll 1$ and there is no adsorption, resulting in
\begin{align}\label{advdiffNDleadcarrier}
    \frac{\partial (\hat u^{(0)} \hat c_N^{(0)})}{\partial \hat x}=0 \, .
\end{align}
Integrating, subject to $\hat u^{(0)} = \hat c_N^{(0)} = 1$ determines (rather obviously)
\begin{align}\label{advdiffNDleadcarrierint}
    \hat u^{(0)} \hat c_N^{(0)}=1 \, .
\end{align}
If we now replace $\hat u^{(0)}$ with equation \eqref{nondimu}, and $\hat c_N$ with equation \eqref{idealgasEoSNDredux} we get
\begin{align}\label{puredux}
    -\frac{\hat{R}^2\hat p^{(0)}}{8\hat p_0}\frac{\partial \hat p^{(0)}}{\partial \hat x}=1\,.
\end{align}
Integrating and applying the boundary condition in \eqref{nondimstokesBC} $\hat p(\hat L,\hat t)=\hat p_L$ we obtain
\begin{align}\label{peq}
    \hat p^{(0)}(\hat x)=\hat p_L\sqrt{1+\left(\frac{\hat p_0^2}{\hat p_L^2}-1\right)\left(1-\frac{\hat x}{\hat L}\right)}\, .
\end{align}
Note that this defines the pressure and hence the velocity as time-independent functions, i.e. $\hat p^{(0)}(\hat x)$ and $\hat u^{(0)}(\hat x)$.

In order to provide a simpler definition of the dimensionless pressure, we rescale pressure as $\bar p=p/p_0$. With the new scaling, equation \eqref{peq} reads
\begin{align}
    \bar p^{(0)}=\sqrt{1-\left(1-\bar p_L^2\right)\hat x/\hat L}\,, \label{p0eqrescale}
\end{align}
and then 
\begin{align}
    \hat u^{(0)}=\mathcal{D}a^{-1}\left(1-\bar p_L^2\right)/\left(\bar p_L\bar p^{(0)}\right)\,,\label{ueqrescale}
\end{align}
where $\mathcal{D}a=16\mu u_0 L/(R^2p_L)$ is the Darcy number.
With the new scaling the pressure at the outlet may be written in terms of the Darcy number
\begin{align}
    1/\bar p_L=\left(\mathcal{D}a/2\right)\left(1+\sqrt{1+4\mathcal{D}a^{-2}}\right)\,.
\end{align}
and then  $\hat u=1/\bar p$.

\bibliography{Main}
\bibliographystyle{vancouver}

\end{document}